\documentclass[%
superscriptaddress,
nofootinbib,
amsmath,amssymb,
aps,
pra,
showkeys,
notitlepage
]{revtex4-1}
\setcitestyle{authoryear,round}

\usepackage{graphicx}
\usepackage{dcolumn}

\usepackage[left=2.5cm,top=3cm,right=2.5cm,bottom=3cm,bindingoffset=0.5cm]{geometry}
\usepackage{graphicx}
\usepackage{hyperref}
\usepackage[bitstream-charter, greekfamily=default]{mathdesign}
\usepackage{lipsum}
\usepackage{xcolor}
\usepackage[utf8]{inputenc}
\usepackage{natbib}

\renewcommand{\epsilon}{\varepsilon}

\graphicspath{{figures/}}

\DeclareMathOperator{\Cor}{Cor}

\newcommand{\avg}[1]{\mathbb{E}\left[ #1\right]}

\newcommand{\dint}{\mathrm{d}}

\begin{document}
\title{From Ants to Fishing Vessels:  A Simple Model for Herding  and Exploitation of \\ \hspace{-0.43cm} Finite~Resources}

\author{Jos\'e Moran}
\email{jose.moran@maths.ox.ac.uk}
\affiliation{Chair of Econophysics and Complex Systems, Ecole polytechnique, 91128 Palaiseau Cedex, France}
\affiliation{Centre d'Analyse et de Math\'{e}matique Sociales, EHESS, 54 Boulevard Raspail, 75006 Paris, France}
\affiliation{LadHyX UMR CNRS 7646, Ecole polytechnique, 91128 Palaiseau Cedex, France}
\affiliation{Mathematical Institute and Institute for New Economic Thinking at the Oxford Martin School, University of Oxford, Oxford, United Kingdom}
\affiliation{Complexity Science Hub Vienna, Josefst\"adter Stra{\ss}e 39, A-1080, Austria}

\author{Antoine Fosset}%
\affiliation{Chair of Econophysics and Complex Systems, Ecole polytechnique, 91128 Palaiseau Cedex, France}
\affiliation{LadHyX UMR CNRS 7646, Ecole polytechnique, 91128 Palaiseau Cedex, France}

\author{Alan Kirman}
\affiliation{Centre d'Analyse et de Math\'{e}matique Sociales, EHESS, 54 Boulevard Raspail, 75006 Paris, France}

\author{Michael Benzaquen}%
\affiliation{Chair of Econophysics and Complex Systems, Ecole polytechnique, 91128 Palaiseau Cedex, France}
\affiliation{LadHyX UMR CNRS 7646, Ecole polytechnique, 91128 Palaiseau Cedex, France}
\affiliation{Capital Fund Management, 23 Rue de l'Universit\'{e}, 75007 Paris, France \medskip}

\begin{abstract}
We analyse the dynamics of fishing vessels with different home ports in an area where these vessels, in choosing where to fish, are influenced by their own experience in the past and by their current observation of the locations of other vessels in the fleet. Empirical data from the boats near Ancona and Pescara shows stylized statistical properties that are reminiscent of Kirman and Föllmer's ant recruitment model, although with two ant colonies represented by the two ports. From the point of view of a fisherman, the two fishing areas are not equally attractive, and he tends to prefer the one closer to where he is based.  This piece of evidence led us to extend the original ants model to a situation with two asymmetric zones and finite resources. We show that, in the mean-field regime, our model exhibits the same properties as the empirical data. We obtain a phase diagram that separates high and low herding regimes, but also fish population extinction. Our analysis has interesting policy implications for the ecology of fishing areas. It also suggests that herding behaviour here, just as in financial markets, will lead to significant fluctuations in the amount of fish landed, as the boat concentration on one area at a given point in time will diminish the overall catch, such loss not being compensated by the reproduction of fish in the other area. In other terms, individually rational behaviour will not lead to collectively optimal results.
\end{abstract}

\date{\today}
\maketitle

\setlength{\parskip}{\medskipamount}
\section{Introduction}

A problem of general interest is that of the individual and collective exploitation of a resource. Depending on the particular context, the dynamics can be very different. A crucial factor is the effect of the behaviour of individuals on the collective outcome. In financial markets for example, the decision to buy may enhance the value of the resource for others as the price of an asset may increase as the demand for it grows. This positive feedback can lead to ``herd behaviour'' and to creation of ``bubbles''. If, on the other hand, the resource is in fixed supply or can only generate a limited flow, as in the case of agricultural production, over exploitation can lead to its exhaustion when individuals do not take account of the overall consequences of their actions. This leads to what has been called ``The Tragedy of the Commons'' in~\citet{hardin1968tragedy}.  

In this paper we  use a version of a model which was developed in the context of financial markets but we modify it to look at a problem of exhaustible resources, in particular that of fisheries. There is a substantial literature on fishing management which analyses the causes of over exploitation and the behaviour that leads to this. Much of that literature was based on understanding the strategies that individual boats use to decide when and where to fish. The simplest idea is that the individuals base their decisions on Catch per Unit Effort (CPUE), see~\cite{gavaris1980use}. This suggests that boats fish until their catch falls below a certain threshold and then move on. This is a purely individualistic model and argues that past individual experience is an adequate basis for decision making. Two questions arise here. Firstly, can one deduce the aggregate behaviour from the observed behaviour of individual vessels, and secondly, does the behaviour of other vessels influence the choices of a particular boat? The answer to the first question lies in the  development of satellite technology which allows individual vessels to be identified and followed; this information provides a basis for analysing the individual and collective behaviour of fishing fleets.  It is, of course, known that vessels do not act in total isolation and a model using tracking data for New Zealand fisheries was, for example, studied in ~\cite{vignaux1996analysis}. This came to the conclusion that “there is evidence that vessels make decisions about where to fish based on both their own recent catch history and on observation about the location and aggregation of other vessels. There is no evidence that there is enough information transfer for vessels to make decisions on the basis of catch rates of the other vessels in the fleet”. What was suggested was that while the influence of other players is taken into account, because of the limited information about the performance of other vessels it may not be the major driving influence for collective behaviour. 

However, a more radical approach, abandoning  simple optimization had been developed earlier by~\cite{allen1986dynamics}. They developed models in part based on the Lotka-Volterra equations which already incorporated recent advances in the understanding of the evolution of complex systems. They studied herd behaviour and simulations of a dynamic model of a Nova Scotia fishery. Their analysis revealed that human responses amplify rapid random fluctuations in recruitment and excite strong Lotka-Volterra type oscillations in a system that would normally settle to a stable stationary state. Their dynamic, multi-species, multi-fleet spatial model was calibrated to the Nova-Scotian groundfish fisheries. They examined the role of ``exploration'' and ``exploitation''. They identified two types of hunters, ``stochasts'' or high-risk takers, and ``Cartesian'' followers, or low risk takers. The result of the interaction between the two reveals, as they say, ``the `out of phase' relationship between abundance and the ease with which fishermen locate a highly sought species and its converse''. They emphasize, contrary to more conventional analysis, ``the importance of information exchange in defining the attractivity of a particular fishing zone to different fleets and the ability of the model to take into account coded information, misinformation, spying and lying; and the fact that models based on global principles, such as `optimal efficiency' or `maximum profit', are clearly of dubious relevance to the real world.''  The crucial difference between this and the work previously cited is that much more weight is given to information about the activity of others and the content of the messages about that activity is assumed to be much richer. 

Our approach is in this spirit and is based on a model in which agents are ``recruited'' to a source of profit by those already benefiting from that source. The actors follow simple rules but their interaction can produce interesting dynamics. A related approach by computer scientists in~\cite{dascalu2013using} suggested that the result might be that of a uniform distribution across the space in which the resource is found. We show that, depending on the weight given to the behaviour of others, vessels can typically operate near to their home port with occasional excursions to another area, but that changing the parameters of the model can lead to a persistent mixing of the two fleets with some boats from each area fishing in the other area. Since what is important is the probability that a boat follows others, the distribution of the boats over the two areas is determined by a stochastic process. This recalls a result of Allen and McGlade in which the survival of the fishery was dependent on the existence of some vessels which chose the place to fish at random and, as in many models of interaction, a degree of randomness may make an important contribution to the overall dynamics of the system. 

The interplay between the ``Cartesian'' and ``stochast'' type of hunters described by~\cite{allen1986dynamics} is strongly reminiscent of models that study financial markets under the assumption that there is imitation between the agents, as in the work of~\cite{chiarella1992dynamics} and~\cite{lux1995herd}.  An interesting feature of the case that we examine is that we can actually observe the activity of the individual boats and to what extent they follow each other and our model provides an explanation for this. In financial markets it is not, in general, possible to monitor the activity of individual investors and one can only infer the motivation of individuals from the aggregate data. The agreement of our model with data is a confirmation of the ubiquity of imitative behaviour in the decision-making process of humans with access to limited information.

The paper is organised as follows. In Section~\ref{section:empirical}, we present an overview of the data used in this article and introduce all relevant definitions.  Section~\ref{sec:simple_model} introduces a model intended to reproduce the main stylized facts present in the data. In particular, we explicitly show how to write the corresponding Fokker-Planck or Kolmogorov forward equation for this model, and later use stochastic calculus techniques to predict features of the model in the form of higher-order correlators that are later validated by the data. The model is developed in Section~\ref{section:mean-field} using simplifying assumptions that are justified by numerical simulations. These allow us to solve for the stationary distribution of the Fokker-Planck equation. Finally, in Sections~\ref{sec:sym_limit} and ~\ref{sec:exploitation_of_finite_resources}, we discuss further consequences of the model and in particular the different collective ``phases'' that describe the aggregate behaviour of fishing vessels. We find that herding which may be rational from the individual point of view, has a detrimental effect on the overall yield of the areas recalling the ``Tragedy of the Commons''.

\section{Empirical fishing data}\label{section:empirical}

As mentioned above, while applicable to a wider range of situations, our work was originally inspired by imitation and herding effects in fishing areas. Here we present the data we use together with some stylized facts, both quantitative and qualitative.

\subsection{Description of the data}

We use the Fishing Vessels Dataset from Global Fishing Watch~\cite{GFWvessels} from Octobre 2012 to December 2016. Since our aim is to analyse the behaviour of fishermen seeking to exploit clearly distinguishable fishing areas, we geographically focus on the Adriatic Sea and specifically on the area encompassing the Italian cities of Ancona and Pescara in which two of the largest fishing harbours and fish markets are situated (see for example the work of ~\cite{gallegati2011s} for a detailed study and description of the Ancona fishing market).\footnote{Note that with this publicly available database, our analysis can be reproduced in any other place of the world where two competing harbours lie reasonably close to each other.} The two cities are separated by a reasonable distance of about $150~\text{km}$, meaning that boats based in one city can easily find themselves fishing close to the other. Further, the existence of large and comparable fish markets in both cities hints at the possibility of matching fishing activity to market data, provided of course one has access to the latter. Note that while another city, San Benedetto del Tronto, lies between Ancona and Pescara, it is responsible for a rather negligible amount of the activity in the area.

We have also restricted our analysis to the behaviour of trawlers. These boats have a low cruise speed and fish in shallow waters close to the coast.  A reasonable hypothesis, which we have confirmed with the local market authorities, is that trawlers fishing in the area are based in either one of the two cities and go out for a short amount of time before coming back to sell their catch on the local market. In particular we were told that, due to the policy of the market to sell fresh local fish, vessels (almost) always get back to the port after 24 hours. We were also told that while there is no ban for a boat registered in a given port to land their fish elsewhere, this seldom happens.\footnote{According to the director of the Ancona fish market, there are no relationships with nearby wholesale markets (Pescara and San Benedetto del Tronto), and, two or three times a year, it happens that a boat based in the nearby port in the north (Fano or Cattolica) comes to sell.} In other words, one expects trawlers based in, say, Pescara to leave port, fish for a day and then come back to sell their catch. 

The reduced data set consists of daily tracking of these trawlers, identified by their 9-digit Maritime Mobile Service Identity (MMSI)  number. Each vessel is tracked on a latitude-longitude grid with resolution $0.1\time0.1$ squared degrees. At Ancona and Pescara's latitude ($\approx 43^{\circ}$ North), this implies a spatial resolution of $\approx 11\times8~\text{km}^2$ (latitude by longitude). Finally, a preliminary study of the data shows that there is a significant reduction of fishing activity from Friday to Sunday, consistent with markets being open Monday through Thursday only. We have thus dropped the former from our data set, keeping only trading days to ensure significant fishing activity.

\begin{figure}[t!]
  \centering
  \includegraphics[width=\columnwidth]{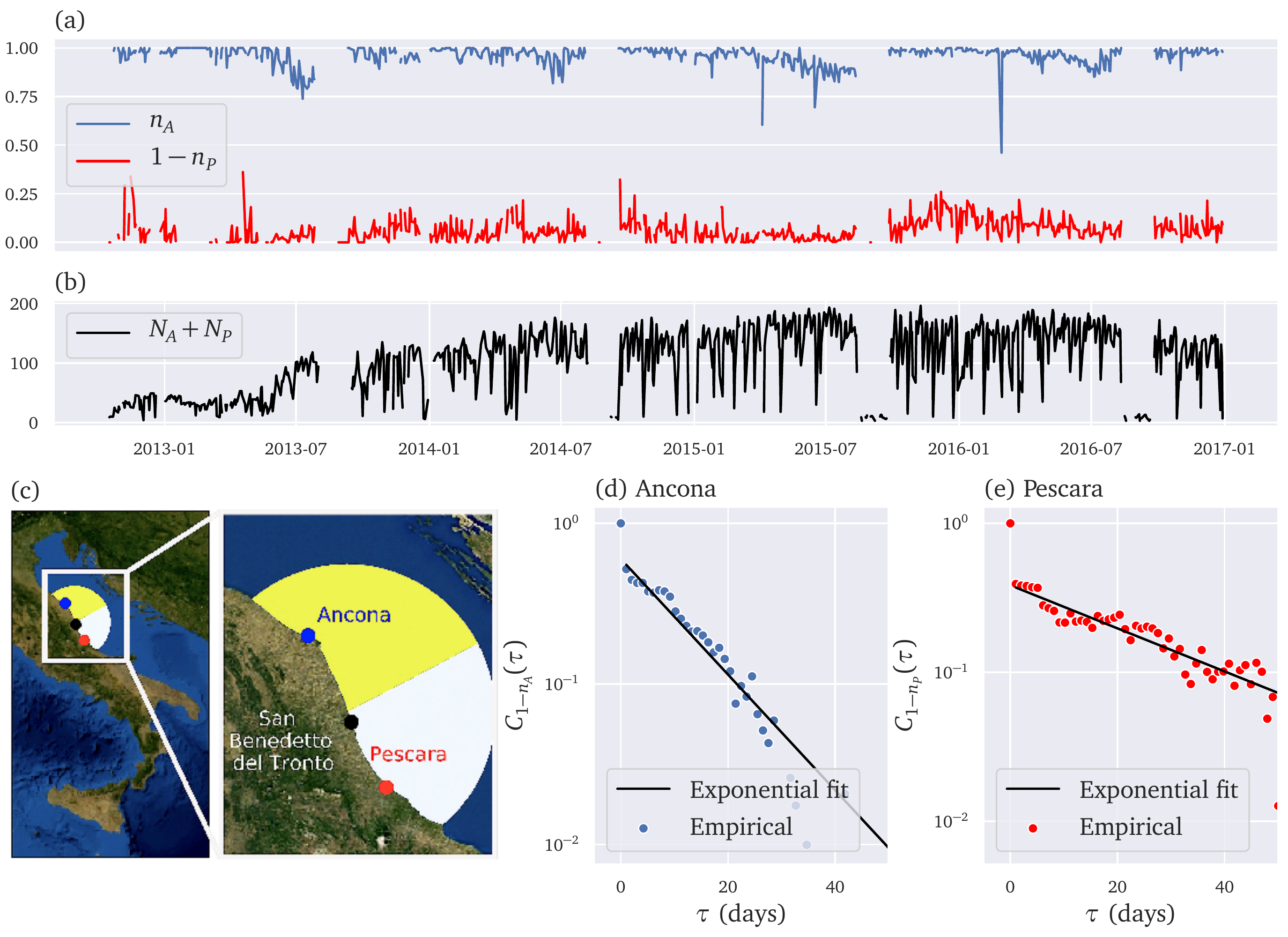}
  \caption{Description of empirical data. Blue curves and markers correspond to data related to the area of Ancona, while red curves and markers correspond to Pescara. (a) Plot of the fractions $n_i(t)$, as defined in Eq.~\eqref{eq:n_def} (b) Plot of the total number of active boats through time $N_A+N_P$. (c) Satellite view of the Adriatic Sea along with the areas $\mathcal{D}_{\mathrm{A}}$ and $\mathcal{D}_{\mathrm{P}}$, as  defined in Eq.~\eqref{eq:areas}. (d) and (e) Autocorrelation functions $C_{1-n_i}(\tau)$, as defined in Eq.~\eqref{eq:corr_function} for both zones. For Ancona we find an exponential fit with a decay rate of $\approx 11 $ days, while for Pescara we find a decay of $\approx 33$ days.}
  \label{fig:fishing_zones_stats}
\end{figure}

\subsection{Defining fishing areas}

To assign each trawler to its base port (Ancona or Pescara), we use 
the following heuristic procedure, which we then cross-validate with MMSI data provided by the Ancona market authorities.  We introduce the notations:\vspace{-0.01cm}
\begin{itemize}\setlength\itemsep{-0.1em}
    \item $h^i(x,t)$ the time spent by trawler $i$ fishing at grid-point $x$ on day $t$,
    \item $w^i(x) := \sum_t h^i(x,t) / \sum_{y,s} h^i(y,s)$, for the average fraction of time spent by trawler $i$ fishing at point $x$,
    \item $d_{\mathrm{A}}(x) $ the distance between point $x$ and Ancona, and $d$ the distance between the two cities,
    \item $d^i_{\mathrm{A}} := \sum_x w^i(x)d_{\mathrm{A}}(x) $, the average distance separating trawler $i$ and Ancona when it is fishing,
    \item ${D_{\mathrm{A}}^i} := \sum_x w^i(x) [d_{\mathrm{A}}(x)]^2  $, the average square distance between trawler $i$ and Ancona,
\end{itemize}{}
and of course symmetrically for Pescara with index P. We then define the neighbourhood of Ancona and Pescara as the pseudo-ellipsoid with focal points the two ports, i.e. the set $\{x~|~d_{\mathrm{A}}(x)^2+d_{\mathrm P}(x)^2\leq 2d^2\}$, of course excluding land, see Fig.~\ref{fig:fishing_zones_stats}(c). We restrict our analysis to trawlers evolving within this area, namely $\{i~|~D_{\mathrm{A}}^i+D_{\mathrm P}^i \leq 2d^2\}$.
We then assign the trawlers to one of the two ports according to their average distance to each of them.
 Defining two distinct areas as: 
\begin{subequations}\label{eq:areas}
\begin{align}
\mathcal{D}_{\mathrm{A}} &= \left\{x~|~ d_{\mathrm{A}}(x) \leq d_{\mathrm P}(x) \quad\text{and}\quad d_{\mathrm{A}}(x)^2 + d_{\mathrm P}(x)^2 \leq 2 d^2\right\}\\
\mathcal{D}_{\mathrm P} &= \left\{x~|~ d_{\mathrm P}(x) < d_{\mathrm{A}}(x) \quad\text{and}\quad d_{\mathrm{A}}(x)^2 + d_{\mathrm P}(x)^2 \leq 2 d^2\right\},
\end{align}
\end{subequations}
a given trawler is assigned to, say, Pescara if its fishing time-weighted average position lies in $\mathcal{D}_{\mathrm P}$. In other words $i\in \text{ Pescara (resp. Ancona)} \text{ if }d_{\mathrm P}^i\leq d_{\mathrm{A}}^i \text{ (resp. } d_{\mathrm{A}}^i< d_{\mathrm P}^i)$. 
 To validate our method of home port identification, we were able confront our classification to the list of the Ancona-based trawlers, kindly provided by the Ancona fish market authorities. Up to a few minor errors, notably related to having identified as Ancona-based a few vessels based in the much smaller San Benedetto del Tronto, the cross-check was successful. Over the whole period we counted $N_\mathrm{A} = 108 $ Ancona-based trawlers and $N_\mathrm{P}=118$ Pescara-based trawlers.

\subsection{Stylized facts}
\label{sub:stylized_facts}

Having tagged each boat to either Ancona or Pescara, we now turn to studying the dynamics of fishing within the two areas $\mathcal{D}_{\mathrm{A}}$ and $\mathcal{D}_{\mathrm P}$. We define the fraction $n_{\mathrm{A}}(t)$ of time spent by Ancona-based  vessels fishing in $\mathcal{D}_{\mathrm{A}}$  namely:
\begin{equation}\label{eq:n_def}
n_{\mathrm{A}}(t) = \frac{\sum_{x\in\mathcal{D}_{\mathrm{A}},i\in\text{Ancona}}h^i(x,t)}{\sum_{y,i\in\text{Ancona}}h^i(y,t)},
\end{equation}
and vice-versa $n_{\mathrm P}(t)$ for Pescara.  Figure~\ref{fig:fishing_zones_stats}(b) displays the evolution of $n_{\mathrm{A}}(t)$ and $n_{\mathrm B}(t)$ throughout the period of interest. While these fractions are most often very close to 1, indicating as one would intuitively expect that trawlers spend most of their time fishing near their home port, one can see, however, that they regularly undergo persistent excursions, revealing that a sizeable fraction of the vessels in each area decide collectively to go elsewhere.\\

\begin{figure}[tb]
  \centering
  \includegraphics[width=\textwidth]{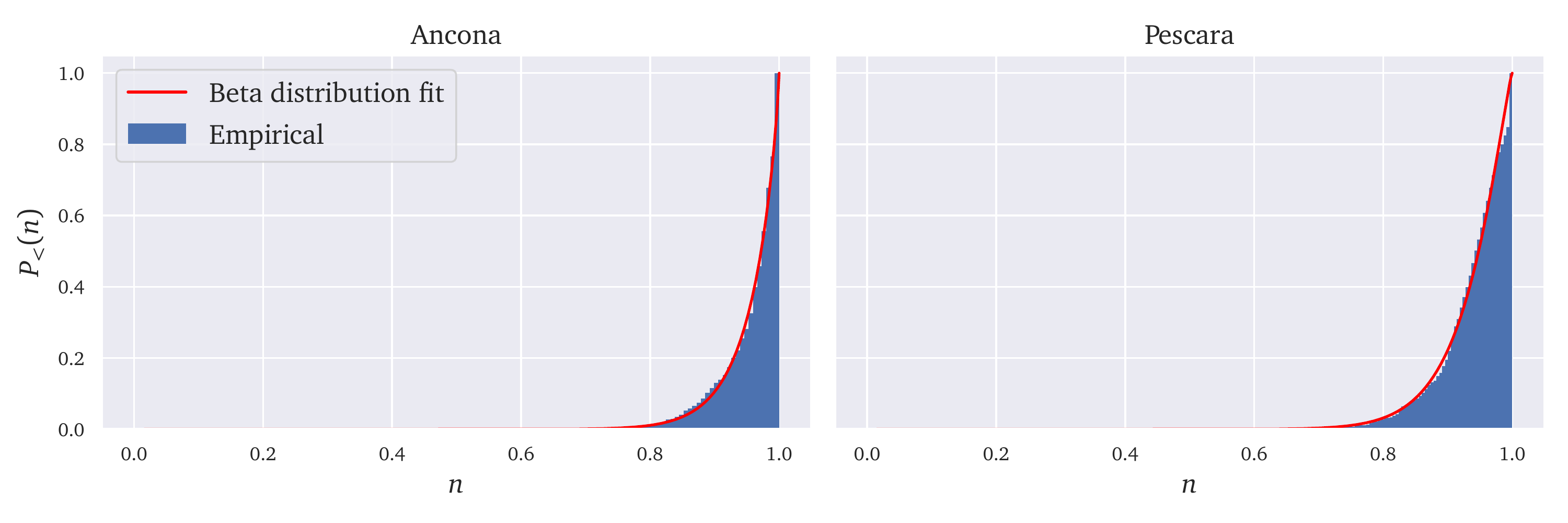}
  \caption{Cumulative distribution function (cdf) of the fractions $n_{\mathrm{A}}$ and $n_{\mathrm{P}}$ as defined in Eq.~\eqref{eq:n_def}. The solid red curves correspond to a fit with a generalized Beta distribution, which has a cdf given by $P_>(n)= C\int_0^n\dint x~ x^{\gamma_0 - 1} (1-x)^{\gamma_1 - 1} $ with $C$ a normalization constant. The parameters for Ancona are $ \gamma_0 = 18.48$ and $ \gamma_1 = 0.82$, while those for Pescara read $ \gamma_0 = 17.73$ and $ \gamma_1 = 1.27$.}
  \label{fig:cdf_data}
\end{figure}

To evaluate the typical length of such excursions, Figs.~\ref{fig:fishing_zones_stats}~(d) and (e) display the auto-correlation functions: 
\begin{equation}\label{eq:corr_function}
C_{1-n}(\tau):=\Cor\left(1-n(t+\tau),1-n(t)\right),
\end{equation} 
for both $n_\mathrm{A}(t)$ and $n_\mathrm{P}(t)$. These are well fitted by the sum of a delta-peak at $0$, which can be attributed to measurement noise and other exogenous factors such as the weather,  
and an exponentially decaying function with typical time-scale ranging from $\approx 11$ to $\approx 30$ days. 
Interestingly enough, Fig.~\ref{fig:cdf_data} reveals that the empirical  distributions of $n_{\mathrm{A}}$ and $n_{\mathrm P}$ are remarkably well fitted by a Beta distribution. 
This is exactly what one obtains in Kirman and Föllmer's ant recruitment model defined in~\cite{kirman1993ants} and \cite{moran2020schr}, in which the Beta distribution emerges as the stationary distribution describing a colony of ants preying on two distinct food sources. Such a distribution  also emerges as the stationary distribution describing genetic populations between two competing alleles~\cite{moran_1958,wright1942}. The key ingredient in these models is the competition between two different entities, be they food sources or genetic alleles. In Kirman and Föllmer's ant model however, the two food sources are strictly equivalent and the resulting Beta distribution describing the fraction of ants at each source is necessarily symmetric, at odds with the results obtained in the present setting. This motivates the asymmetric zones model introduced below. Another significant difference with Kirman and Föllmer's original model is that the "food sources" here are not inexhaustible, as fish cannot  reproduce at an infinite rate.

These empirical results and observations motivate us to introduce a model extending Kirman and Föllmer's original ant recruitment model to our context. In essence, one can think of the two cities as two distinct ant colonies that can obtain their food from either of the two zones. For each colony, the further fishing area is necessarily less attractive, leading to the asymmetric character of the distribution. In addition, not being in a setting with unlimited resources, our model should take into account the fact that over-fishing may deplete the sea.

\section{A simple model}\label{sec:simple_model}

Kirman and Föllmer's original ant-recruitment model~\cite{kirman1993ants} was successful at explaining a rather puzzling fact well known to entomologists~\cite{deneubourg1990self, beckers1990collective}. Ants, faced with two identical and inexhaustible food sources tend to concentrate on one of them and occasionally switch to the other. In the model, at each time step a given ant may either (i) encounter another ant from the other inexhaustible food source and decide to switch to her peer's source (be recruited), or (ii) spontaneously decide to switch food sources without interacting. The driving mechanism of the dynamics results from the trade-off between the intensity of the noise-term $\varepsilon$ (spontaneous switching), and that of the interaction term  $\mu$, see also~\cite{moran2020schr}.

Here we present an extension of Kirman and Föllmer's original model to account for \emph{exhaustible} and asymmetric sources, notably aimed at accounting for some of the stylized facts presented in the previous section for fishing areas.  Seeking to model fishermen exploiting a set of fishing areas, we imagine that boats follow the same basic dynamics as the ants: if they initially fish within a certain zone, they may decide to move elsewhere either because they see their peers fishing there, deciding to imitate them because they assume that their yield is good, or spontaneously decide to move elsewhere randomly for the sake of exploration.

As briefly mentioned above, this model is mathematically equivalent to many models appearing in different fields. It was originally proposed in~\cite{moran_1958} as a way to model the population dynamics between two competing alleles of the same gene, A and B; the equivalent of the imitation between ants is in this case the self-reinforcing reproduction of each allele, while the spontaneous switching corresponds to a random mutation from A to B and vice-versa. The extension of the ant model that is most similar to ours is the one described in~\cite{alfarano005}, where the authors study a model that is equivalent to having two {\it asymmetric} food sources, meaning that one is more attractive than the other, as an analogy to model herding and volatility bursts in financial markets. The same idea of using herding to explain fat tails in the distribution of financial returns was also exploited by~\cite{sano2015}, although without the introduction of asymmetry. As we shall see, similar behaviour in our context can lead to increased volatility in both the fish population and the amount of fish that is caught.

It has also been used in the social sciences within the so-called noisy voter model studied in~\cite{Carro2016} and is closely linked to other voter models (see e.g. ~\cite{Redner2019} for a recent review). In this setting the ants become voters that have to choose between two competing political choices, and make their pick because of some idiosyncratic bias (akin to the spontaneous switching) or because they are influenced by their peers.

There is a key difference, however, between the ant model and some instances of the voter model: it may be that the mapping between the two models is exact, and at each time step a voter can pick one of their neighbours and imitate their choice, leading to exactly the same dynamics as for the ant model. Another version of the model, however, can instead imagine that the voter has again two different choices, but polls his peers and decides to adopt the choice of the majority. This second version leads to different dynamics, where multiple equilibria appear and correspond to states where the agents self-reinforce their choice, and may collectively change their mind after an exogenous shock, however small, as it happens with the random field Ising or random utilities model described in~\cite{Galam_Shapir_1982,Gordon_Nadal_Phan_Vannimenus_2005,Nadal_Phan_Gordon_Vannimenus_2003,Michard_Bouchaud_2005,Bouchaud_2013}. These models often lead to a very spectacular phenomenology of ``avalanches'' and exhibit true phase transitions, along with hysteresis, see e.g.~\cite{Kuntz1999}.

This family of models has also gained traction in the finance and economics community, as a way to understand the origins of large endogenous fluctuations in financial markets~\cite{lux1995herd}, whose origin may be found in the word-of-mouth way that information diffuses among market participants~\cite{Shiller1989}. In fact, the model can be embedded into an equilibrium asset pricing model, as in~\cite{alfarano_2008}, and the dynamics one obtains are strikingly similar to models of financial markets with two types of agents -- fundamentalists who attempt to find the true, ``fundamental'' price of an asset and followers who imitate them. Good examples of these models can be found in e.g.~\cite{chiarella1992dynamics,alfarano_2008}, and recent empirical confirmation of the main qualitative findings of these models is given in~\cite{Majewski2018,bouchaud2017black}.

As discussed above, our model has two major differences that depart from the original ant-recruitment model and its siblings. First, we consider that a fishing area has finite resources: fish reproduce until reaching a certain finite capacity but they are also depleted by fishermen in the area (as in e.g. MacArthur's models  described in \cite{macarthur1970species, mac1969species}). As a consequence, we assume that the random switching rate at which fishermen decide to depart from a given area depends on the fish population of that area. Note that this is very close in spirit to the modelling done in ~\cite{allen1986dynamics}, albeit that our model takes into account imitative behaviour in fishermen. The second difference with the ant model is that we imagine two ``colonies'' instead of just one, corresponding to vessels based at the two different fishing ports of Ancona and Pescara. Guided by the idea that fishermen prefer to go to areas close to their own home port, notably to save fuel, we introduce an asymmetry between the fishing areas for each food source. 

The two ports, labelled $\mathrm{A}$ and $\mathrm{P}$, have two distinct populations of fishermen, which may decide to exploit two fishing areas, $S_1$ and $S_2$, with the fishermen from $\mathrm{A}$ preferring to fish at $S_1$ and vice versa. One may of course reasonably argue that this view is far too coarse-grained, and that there may be, for example, many different fishing areas that are available close to each port. It is however possible to show under mild hypotheses that the two zones $S_1$ and $S_2$ in the model can be seen as the aggregation of a large number of smaller areas, with the same dynamics, see Appendix \ref{appendix:multi_zones} for details. For clarity, we shall define the model in discrete time, before moving into continuous time for analytical convenience.

Without loss of generality, we focus only on the dynamics of fishing vessels at one of the two ports, say Ancona, as we assume that fishermen only interact with boats coming from the same city.\footnote{Anecdotal evidence suggests indeed that the main interaction between people working in different boats happens at port in the fishing market or during informal conversation.}
We define now $N_A$ and $N_P$ as the number of boats based at Ancona and Pescara respectively, and let each of them decide to go to any of the two areas $S_1$ and $S_2$. We denote $m_i(t)$, with $i=1,2$, their respective fish populations at time $t$, and further assume that:
\begin{itemize}  \setlength\itemsep{-0.2em}
    \item Boats only fish in one area each day and come back to that area if they do not decide to switch to another one for the next day.
    \item A vessel's catch $c_i(t)$ is proportional to the amount of fish available in the area: $c_i(t) = \frac{\beta}{N_\mathrm{A}} m_i(t)$ with $\beta/N_\mathrm{A}\in[0;1]$.\footnote{Without changing our main conclusions, one could also allow for noise by drawing $c_i(t)$ from a given distribution centred about $\beta m_i(t)/N$. This would allow us to introduce randomness into the fishing efficiency of each trawler, an interesting extension that we leave for further work.}
    \item Fish reproduce at a multiplicative rate $\nu_i$, which we take to be equal to $\nu$ for both areas.
    \item As a first approximation, fish do not travel from one area to the other.\footnote{This constraint can be easily relaxed by e.g. adding a migration term where fish from $2$ move to $1$ at a certain rate and vice-versa. In practice, this would only tend to prevent the difference between the two fish populations from fluctuating too wildly.}
    \item The fish population within any area cannot exceed a carrying capacity $K_i$, which is the maximal population that can be present within an area in the absence of fishing. This carrying capacity is the same for all areas, as we have taken all of them to be equivalent. Without loss of generality, we take $K_1=K_2=1$ in all that follows. 
\end{itemize}

Note that these definitions, which also amount to thinking of the fish population as consisting of the same species in both areas, are partially justified by our considering only trawlers, that therefore fish only very specific, shallow water dwelling species. 

We further define $N_{\mathrm{A},i}(t)$ the number of vessels from port $\mathrm{A}$ fishing at zone $i$ at time $t$ (and $N_{\mathrm P,i}(t)$ respectively). The number of fishing vessels in each port is fixed, implying for all $t$: $N_{\mathrm{A},1}(t)+N_{\mathrm{A},2}(t) = N_\mathrm{A}$. Our assumptions translate into the following continuous-time evolution for the fish population:
\begin{equation}\label{eq:fish_pop}
    \frac{\mathrm{d}m_i}{\mathrm{d}t} = m_i(t) \left[\nu g(m_i(t)) - \frac{\beta}{N_{\mathrm{A}}} \left(N_{\mathrm A,i}(t)+N_{\mathrm P,i}(t)\right)\right],
\end{equation}
where the function $g$  must satisfy $g(0)=1$ and $g(1) = 0$.

On the right hand side, the first term $\nu m_i g(m_i(t))$ models the amount of fish that spawned between $t$ and $t+\mathrm{d}t$ in absence of fishing; it is given by a birth-rate $\nu$ when the fish are not too plentiful, but goes to $0$ as the zone gets saturated and cannot sustain any more fish. The simplest assumption one can make is that of logistic growth, which is the most common assumption used to model the dynamics of ecosystems, leading to $g(m_i(t)) = 1-m_i(t)$. It follows that $m_i(t)\in[0;1],\ \forall t$, where $m=1$ corresponds to a fishing area at full capacity and $m=0$ corresponds to a depleted area.

Under these assumptions, the evolution of the fish population is of the Lotka-Volterra type, as advocated in~\cite{allen1986dynamics}.\footnote{As an interesting anecdote, we learned in \cite{Bacaer_2011} that ``Vito Volterra was born in the Jewish ghetto of Ancona in 1860, shortly before the unification of Italy, when the city still belonged to the Papal States'', and that ``in 1925, at age 65, Volterra became interested in a study by the zoologist Umberto D’Ancona, who would later become his son-in-law, on the proportion of cartilaginous fish (such as sharks and rays) landed in the fishery during the years 1905–1923 in three harbours of the Adriatic Sea: Trieste, Fiume and Venice. D’Ancona had noticed that the proportion of these fish had increased during the First World War, when the fishing effort had been reduced''. This led him to take interest in models that Alfred Lotka had first used to model very general population dynamics, and that we now apply, without knowing any of this at first, to the fish population dynamics at the ports of Ancona and Pescara.}  The second term on the right hand side corresponds simply to the decrease in fish population because of fishing activity. 

Furthermore, we assume that a fishing vessel based at $\mathrm{A}$ fishing at $i$ can randomly decide to go elsewhere with probability $\varepsilon_{\mathrm{A},i} f(m_i(t))$, where the function $f$ satisfies $f(1)=1$ and $f(0)=1+\kappa$. Here, $\varepsilon_{\mathrm{A},i}$ controls the base intensity of the noise, that can take a maximal value $\varepsilon_{\mathrm{A},i}(1+\kappa)$ when the zone is depleted. The variable $\kappa$ therefore captures the additional percentage by which the spontaneous switching probability is raised when the area is depleted. If $f$ is supposed to be linear, an assumption that we will make below, then $\kappa$ also captures the sensitivity of the spontaneous switching probability with respect to changes in the fish population.

Fishermen therefore have a higher incentive to go elsewhere as their fishing yield decreases, but they still prefer to fish near their home port. We highlight the preference of fishermen from $\mathrm{A}$ for zone $1$ by setting $\varepsilon := \varepsilon_{\mathrm{A},1}=\varepsilon_{\mathrm{A},2}/C_{\varepsilon}$ with $C_{\varepsilon}>1$  a parameter controlling the degree of asymmetry between zones $S_1$ and $S_2$ for a fisher from $\mathrm{A}$. This allows us to have a higher spontaneous switching rate $S_2\rightarrow S_1$ for fishermen from $\mathrm{A}$. 

Besides this random switching rate, we add in the crucial element in our model, which is that agents imitate each other. Each day, a fisherman randomly picks one of his peers and decides to imitate him with probability  $\propto \mu$, so that $\mu$ sets the propensity for imitation in the model. In this case, the probability that a boat from $\mathrm{A}$ initially at zone $S_i$ decides to move to zone $S_j$ is given by:\footnote{Note that this expression takes into account a small typo in~\cite{moran2020schr}, where the imitation probability is proportional to $\mu/N$, while it should be proportional to $\mu N$.}
\begin{eqnarray}\label{eq:vessel_transition}
    P_{\mathrm{A}}(S_i\rightarrow S_j) = \varepsilon_{\mathrm{A},i} f(m_i(t)) + N_{\mathrm{A}}\mu \frac{N_{\mathrm{A},j}(t)}{N_\mathrm{A}-1}.
\end{eqnarray}
Writing then $n_{\mathrm{A},i}=N_{\mathrm{A},i}/N_{\mathrm{A}}$, we take the limit $N_{\mathrm{A}},N_{\mathrm P}\to\infty$ with $N_P/N_\mathrm{A}=C_N$ fixed. Within this limit, we denote $\mathbf{n}_A=(n_{\mathrm{A},1}, n_{\mathrm{A},2})$ and $\mathbf{m}=(m_1,m_2)$ and study the probability density $\rho(\mathbf{n}_\mathrm{A},\mathbf{n}_{\mathrm P}, \mathbf{m})$ of all the variables of the model.

The equation one obtains is called the Fokker-Planck equation~\cite{risken1996fokker}, also known as the Kolmogorov forward equation in applied mathematics and closely related to the Hamilton-Jacobi-Bellman equations describing the optimal choice of a rational agent; it can be interpreted as the continuous limit of a Markovian transition matrix. An explicit example of how to obtain this equation for the ant model, which can be easily extended to the one presented here, is presented in~\cite[Appendix~A]{moran2020schr}. For the sake of pedagogy, we will explicitly work out how to obtain the Fokker-Planck equation for the standard ant model below.

Consider a simplified version of the model, where the noise $\varepsilon$ is constant and where there are only two different areas, analogous to the two food sources for the ants. We thus focus on the dynamics followed by the boats in one port only, and we write $N_{\mathrm{A},1}=k$ and $N_{\mathrm{A}}=N$. In this case, the two possibilities at each step are to have $k\to k+1$ as one boat at zone $2$ goes to zone $1$ or $k\to k-1$ when a boat does the opposite. These two probability transitions read, following Eq.~\eqref{eq:vessel_transition}, 

\begin{equation}\label{eq:transition_probs}
\begin{split}
P(k\to k+1) &= \left(1-\frac{k}{N}\right)\left(\varepsilon + \mu N \frac{k}{N-1} \right)\\
P(k\to k-1) &= \frac{k}{N}\left(C_{\varepsilon}\varepsilon + \mu N \frac{N-k}{N-1} \right).
\end{split}
\end{equation}

The probability equation evolves according to\footnote{To take the continuous limit in time, it is necessary to define the timestep $\dint t:=1/N$, which will go to $0$ as the limit $N\to\infty$ is taken. This explains the factor in front of the time-derivative.}
\begin{equation}\label{eq:master_eq_simple}
\begin{split}
\frac{1}{N}\frac{\partial P(k,t)}{\partial t} = &P(k+1\to k)P(k+1,t) - P(k\to k+1)P(k,t)\\
& + P(k-1\to n)P(k-1,t) - P(k\to k-1)P(k,t).
\end{split}
\end{equation}

Working in the $N\to\infty$ limit however pushes us to introduce the density $\rho(n,t)$ of the variable $n=k/N$. Rewriting in terms of this variable and in the $N\to\infty$ limit, the term e.g. $P(k+1\to k)P(k+1,t)$ becomes:
\begin{equation}\label{eq:ktokp1}
\left(n+\frac{1}{N}\right)\left(\varepsilon+\mu N \left(1-n-\frac{1}{N}\right)\right)\rho\left(n+\frac{1}{N},t\right) \approx  g(n,t)+ \frac{1}{N}\partial_n\left[g(n,t)\right]+\frac{1}{2N^2}\partial_{nn}^2 \left[g(n,t)\right],
\end{equation}
where we've defined $g(n,t)= n\left(\varepsilon+\mu N (1-n)\right)\rho(n,t)$.

Similarly, the term $P(k-1\to k)P(k-1,t)$ involves a similar expression that reads $h(n,t)-\frac{1}{N}\partial_n\left[h(n,t)\right]+\frac{1}{2N^2}\partial^2_{nn}\left[h(n,t)\right]$ with $h(n,t)=(1-n)\left(\varepsilon C_\varepsilon+\mu Nn\right)\rho(n,t)$. Adding all the expressions in Eq.~\eqref{eq:master_eq_simple} and retaining only terms that remain of order $1$ in the limit $N\to\infty$ leads to the Fokker-Planck equation:

\begin{equation}\label{eq:simple_fp}
\partial_t \rho = -\partial_n\left[1-(1+C_{\varepsilon}\varepsilon)n\right]\rho+\mu\partial^2_{nn}\left[n(1-n)\right]\rho.
\end{equation}

For our model, the Fokker-Planck equation is more complicated because it concerns the four coupled quantities described by the vectors $\mathbf{n}_{\mathrm{A}}, \mathbf{n}_{\mathrm{A}}$ and $\mathbf{m}$. The same principle described above holds, and the same type of equation can be derived. Altogether, this equation  reads:
\begin{eqnarray}\label{eq:fokker_planck}
\begin{split}
\partial_t \rho =& - \varepsilon \partial_{n_{\mathrm{A},1}}\left[C_{\varepsilon}f(m_2)-n_{\mathrm{A},1}\left[C_{\varepsilon}f(m_2)+f(m_1)\right]\right]\rho  + \mu \partial^2_{n_{\mathrm{A},1},n_{\mathrm{A},1}}\left[n_{\mathrm{A},1}(1-n_{\mathrm{A},1})\right]\rho \\
& - \varepsilon \partial_{n_{\mathrm{P},2}}\left[C_{\varepsilon}f(m_1)-n_{\mathrm{P},2}\left[C_{\varepsilon}f(m_1)+f(m_2)\right]\right]\rho  + \mu \partial^2_{n_{\mathrm{P},2},n_{\mathrm{P},2}}\left[n_{\mathrm{P},2}(1-n_{\mathrm{P},2})\right]\rho  \\
&- \partial_{m_1}\left[\nu(1-m_1)-\beta \left(n_{\mathrm{A},1}+C_N(1-n_{\mathrm{P},2})\right)\right]m_1\rho\\
&- \partial_{m_2}\left[\nu(1-m_2)-\beta \left(C_Nn_{\mathrm{P},2}+(1-n_{\mathrm{A},1})\right)\right]m_2\rho ,
\end{split}
\end{eqnarray}
where the first two lines govern respectively the time evolution of $n_{\mathrm{A},1}$ and $n_{\mathrm{P},2}$ -- one can recognize in the simple form of these two lines  an expression similar to that of Eq.~\eqref{eq:simple_fp} -- and the last two lines that of the fish population in the two areas.  

Note that keeping the next term of order $1/N$ in Eq.~\eqref{eq:simple_fp} modifies only the diffusive term from $\mu n(1-n)$ to $\mu n(1-n) + \frac{1}{2N}\left(n\varepsilon + (1-n) C_{\varepsilon}\varepsilon\right)$ while leaving the drift term unchanged. Computing the terms of order $1/N_{\mathrm{A}}$ in Eq.~\eqref{eq:fokker_planck} also leads to a modification of the diffusion terms along the same lines. This is the same result as the one obtained in another version of the asymmetric ant model in~\cite{alfarano005}. Note that the analysis done in the Appendix of~\cite{moran2020schr} shows that the finite size effects are negligible in the large $N$ limit, as they only affect values of $n$ in the intervals $[0;1/N]$ and $[1-1/N;1]$.\footnote{Indeed, that Appendix shows that the description is good as long as $\tan~\varphi \ll N$, with $\varphi = \arcsin(2n-1)$. Expanding in $n$ or in $1-n$ shows that this is valid as long as $\frac{1}{N}\ll 1-n$ and $n$.}

These equations fully close the model, which in our view represent the simplest setting for a system with limited resources exploited by entities with a myopic exploration/imitation strategy. As they stand, however, they cannot be solved analytically. We shall now resort to a mean-field approximation to find a solution. 

\section{Mean-field approximation}
\label{section:mean-field}

Solving the Fokker-Planck equation~\eqref{eq:fokker_planck} is a very difficult task, as the different terms that intervene take into account the interactions between the proportion of fishermen in a given zone and the fish population in it. In general, these two quantities fluctuate in time, and the main difficulty lies in unravelling how these fluctuations interact. However, if one is interested in a very aggregated picture, one can simplify the problem significantly by calculating the behaviour of the fish {\it as if} they were only subject to the averaged, fluctuation-free, action of the fishing boats and vice-versa.  This is what is known as the \textit{mean-field approximation}, which allows to replace, as a first step, the behaviour of the fish populations $m_i$  with their long-time averages. 

Note that the nature of this mean-field approximation is different in nature, but not in spirit, to the approximation used to solve similar models with imitation, but where the agent follows the choice of the majority instead of following the choice of a single agent as happens here. In those models, the mean-field approximation consists in saying self-consistently that the opinion of each agent is determined by the opinion of the majority, and that the latter is nothing but the average of the former. Here, the nature of the approximation is different, as we are not replacing the interaction of the agents between themselves with an averaged interaction, but doing it for the interaction between them and the fish instead.

Taking Eq.~\eqref{eq:fish_pop} in the continuous time limit, the evolution of the fish population of, e.g., zone $1$ follows:
\begin{equation}\label{eq:fish_evolution_cont}
    \begin{split}
        \frac{\dint m_1}{\dint t} = m_1(t)\left(\nu (1-m_1(t)) - \beta \left(n_{\mathrm{A},1}+C_N \left(1-n_{\mathrm{P},2}\right)\right)\right).
    \end{split}
\end{equation}
 If we take the average of this equation, we expect the left hand side to be $\frac{\dint \avg{m_1}}{\dint t} =0$ as the average is not expected to fluctuate in time.\footnote{This average is taken with respect to infinitely many possible realizations of the stochastic process describing the evolution of our model. Nonetheless, because the model is ergodic we expect that this averaging is also true when one considers time-averages, meaning that we take a single realization of the process and look at the average of $m_1$ in time.} This yields the following expression for the right hand side:
\begin{equation}\label{eq:mean_field_avg}
\avg{m_1} = \left[1-\frac{\beta}{\nu}\left(\avg{n_{\mathrm{A},1}} + C_N\left(1-\avg{n_{\mathrm{P},2}}\right)\right)\right]_+,
\end{equation}
where $[x]_+ = x\mathbf{1}_{x>0}$ denotes the positive part of $x$. In particular, one can see that there exists a trivial extinction line for the fish population for:
\begin{equation}\label{eq:critical}
\nu=\beta\left[\avg{n_{\mathrm{A},1}} + C_N\left(1-\avg{n_{\mathrm{P},2}}\right)\right],
\end{equation}
which corresponds to the case where the reproductive rate of fish corresponds exactly to the rate at which they are fished. 

We now do the same mean-field approximation the other way around to simplify the evolution of the fishermen. We insert Eq.~\eqref{eq:mean_field_avg} into the vessels' dynamics by replacing the argument of $f(m_i)$ by the average, as $f(\avg{m_i}):=f_i$, which amounts to saying that the boats only interact with the average behaviour of the fish. Choosing, to be precise,  a linear function for $f$, i.e.  $f(x)= 1+\kappa(1-x)$, the average $\avg{n_{\mathrm{A},1}}$ can now be easily computed from Eq.~\eqref{eq:fokker_planck} by setting the drift term to $0$, as:
\begin{equation}\label{eq:avg_na1}
\avg{n_{\mathrm{A},1}} = \frac{C_{\varepsilon}f_2}{C_{\varepsilon}f_2+ f_1}= \frac{C_{\varepsilon}\left(1+\kappa(1-\avg{m_2})\right)}{2+\kappa\left[1-\avg{m_1}+C_{\varepsilon}\left(1-\avg{m_2}\right)\right] }. 
\end{equation}

Therefore, the mean-field approximation applied to the quantities $m_1$ and $n_{\mathrm{A}}$ have allowed to derive the two Eqs.~\eqref{eq:mean_field_avg} and ~\eqref{eq:avg_na1}, and therefore to obtain self-consistently the averages of these quantities. The next step is to obtain a fuller picture of the aggregate behaviour of the vessels within the mean-field approximation, and to obtain for example the probability density associated with it.

\subsection{Stationary solutions} 
\label{sub:solving_the_mean_field_model}
Consistent with our mean-field approximation, we  set $m_1=\avg{m_1}$ (resp. $m_2=\avg{m_2}$) in Eq.~\eqref{eq:fokker_planck} to obtain a Fokker-Planck equation that describes the vessels. Note that the two last lines in that equation, as we have supposed that the fish population reach their stationary values, become $0$. In the previous equation, vessels from the two ports interacted indirectly through fishing in the same zone and depleting the fish in it, motivating all the boats in that zone to leave. Replacing the behaviour of the fish by its average amounts to neglecting this effect, and to an effective decoupling of the two variables $n_{\mathrm{A}}$ and $n_{\mathrm{P}}$. In this case, the Fokker-Planck equation reads

\begin{equation}\label{eq:fp_mf}
\partial_t \rho = \partial_{n_{\mathrm{A},1}} J_1 + \partial_{n_{\mathrm{P},2}} J_2,
\end{equation}
where:
\begin{equation}\label{eq:j1}
J_1 = - \varepsilon \left[C_{\varepsilon}f_2-n_{\mathrm{A},1}\left[C_{\varepsilon}f_2+f_1\right]\right]\rho  + \mu \partial_{n_{\mathrm{A},1}}\left[n_{\mathrm{A},1}(1-n_{\mathrm{A},1})\right]\rho,
\end{equation}
and where the transposition to find the definition of $J_2$ is transparent. The two quantities $J_1$ and $J_2$ are probability fluxes, and for example $J_1\left(n_{\mathrm{A},1},t\right)$ can be interpreted as the probability mass going from $n_{\mathrm{A},1}+\Delta n$ to $n_{\mathrm{A},1}$ for an infinitesimal $\Delta n$ during an infinitesimal amount of time. 

The stationary state is found by setting $J_1=0$ and $J_2=0$, meaning that there is no probability flux in the model, and solving the obtained equations for $\rho$. The decoupling of the two variables $n_{\mathrm{A},1}$ and $n_{\mathrm{P},2}$  allows one to write the density as the product of two independent densities, as
\begin{equation}\label{eq:sol}
\rho(n_{\mathrm{A},1}, n_{\mathrm{P},2}) = \rho_1\left(n_{\mathrm{A},1}\right)\rho_2\left(n_{\mathrm{P},2}\right),
\end{equation}
with:
\begin{equation}\label{eq:solution}
\rho_1(n_{\mathrm{A},1}) = C_1 n_{\mathrm{A},1}^{\gamma_{\mathrm{A},0}-1}\left(1-n_{\mathrm{A},1}\right)^{\gamma_{\mathrm{A},1}-1},\quad \rho_2\left(n_{\mathrm{P},2}\right) = C_2 n_{\mathrm{P},2}^{\gamma_{\mathrm{P},0}-1}\left(1-n_{\mathrm{P},2}\right)^{\gamma_{\mathrm{P},1}-1},
\end{equation}
where $C_1$ and $C_2$ are normalisation constants and the $\gamma$ parameters for the $\rho_1$ distribution (the parameters for $\rho_2$ can be easily deduced) read:
\begin{equation}\label{eq:gamma_params}
\gamma_{\mathrm{A},0} = \frac{\varepsilon}{\mu}C_{\varepsilon}f_2,\quad \gamma_{\mathrm{A},1} = \frac{\varepsilon}{\mu} f_1.
\end{equation}

By direct integration one can express the normalisation constant $C_1$ using the Beta $\mathrm{B}(\alpha,\beta)$, as
\begin{equation}
  C_1 = \frac{1}{\mathrm{B}\left(\gamma_{\mathrm{A,0}},\gamma_{\mathrm{A},1}\right)},\quad \mathrm{B}\left(\alpha,\beta\right):= \int_0^1\mathrm{d}x~x^{\alpha-1}(1-x)^{\beta-1}, 
\end{equation}
and compute the average and variance,
\begin{equation}\label{eq:mean_var}
  \avg{n_{\mathrm{A,1}}}= \frac{\gamma_{\mathrm{A,0}} }{\gamma_{\mathrm{A,0}}+\gamma_{\mathrm{A,1}}},\quad \mathbb{V}\left[n_{\mathrm{A,1}}\right] = \frac{\gamma_{\mathrm{A,0}}\gamma_{\mathrm{A,1}}}{\left(\gamma_{\mathrm{A,0}}+\gamma_{\mathrm{A,1}}\right)^2\left(1+\gamma_{\mathrm{A,0}}+\gamma_{\mathrm{A,1}}\right)},
\end{equation}
in agreement with Eq.~\eqref{eq:avg_na1}.

Note that the mean is independent of the ratio $\varepsilon/\mu$, but that as this ratio is large compared to $1$, the variance scales as $\frac{\mu}{\varepsilon}$, showing that in the high-imitation regime where $\mu>\varepsilon$ an increase in imitation does cause an increase in the volatility of the process.

Note also that full dynamical solutions $\rho(n_{\mathrm{A},1},t),~ \rho(n_{\mathrm{P},2},t)$ can be obtained in terms of hypergeometric functions, in the same spirit as \cite{moran2020schr}, see Appendix~\ref{ap:dynamics}.

The structure of the Beta distribution we obtain for this also allows further justification to the pertinence of the model. It may indeed seem quite arbitrary to split the various possibilities a fisherman has into only two areas, whereas in all generality one may imagine that there exists an arbitrary number $M$ of fishing spots available to him. If we suppose that this is the case, and still consider that there is an interplay between random switching and imitation between agents, then it is possible to study the dynamics of the quantity $n_{\mathrm{A},k}$ which is the fraction of boats from $\mathrm{A}$ fishing in zone $k$, with the condition that $\sum_k n_{\mathrm{A},k}=1$.

The joint stationary distribution of this quantity is found to be a Dirichlet distribution, 

\begin{equation}\label{eq:dirichlet_dist}
\rho(\{n_{\mathrm{A,k}}\})\propto \prod_{k=1}^M \left(n_{\mathrm{A},k}^{\alpha_k}\right)\mathbf{1}_{\left\{\sum_k n_{\mathrm{A},k}^{\alpha_k}=1\right\}},
\end{equation}
where the $\alpha_k$ are exponents that depend on the noise level $\varepsilon_k$ of each area and on the imitation parameter $\mu$. This distribution has a very interesting property, which is that if one splits these $M$ areas into two groups of respectively $m$ and $M-m$ areas by defining $\tilde{n}_1 = \sum_{k=1}^{m} n_{\mathrm{A},k}$ and $1-\tilde{n}_1=\sum_{k=m+1}^{M} n_{\mathrm{A},k}$, then one finds that the stationary distribution of $\tilde{n}_1$ has the same functional form as the one given in Eq.~\eqref{eq:solution}. In other words, our splitting of the dynamics into two areas can be seen, from the perspective of the boats' dynamics, as the aggregation of a multitude of different areas into two. This point is discussed briefly in Appendix~\ref{appendix:multi_zones}.

For the reader unfamiliar with Fokker-Planck equations and stochastic processes, the following thought experiment may help in understanding what we've  stated mathematically above. The model in Sec.~\ref{sec:simple_model} can be run as a computer simulation where $n_{\mathrm{A},1}$ and $n_{\mathrm{P},2}$ correspond to numbers between $0$ and $1$ and can be known at all times. It is then possible to run a very large number of simulations with the same initial conditions for these two variables and to compute cross-sectional histograms for these variables at a given time-point, that is the histograms of the same variable at the same time but through different simulations. The Fokker-Planck equation describes how these histograms will change in time, before eventually settling onto distributions described by Eq.~\eqref{eq:solution}. The model is however ergodic, and if we also take one single simulation and run it for a very long time, we can plot the histogram of, say, $n_{\mathrm{A},1}\left(\{t_i\}\right)$ at randomly sampled times $t_i$ and we will observe a histogram described by the density $\rho_1(n_{\mathrm{A},1})$.

\begin{figure}[tb]
  \centering
  \includegraphics[width=\textwidth]{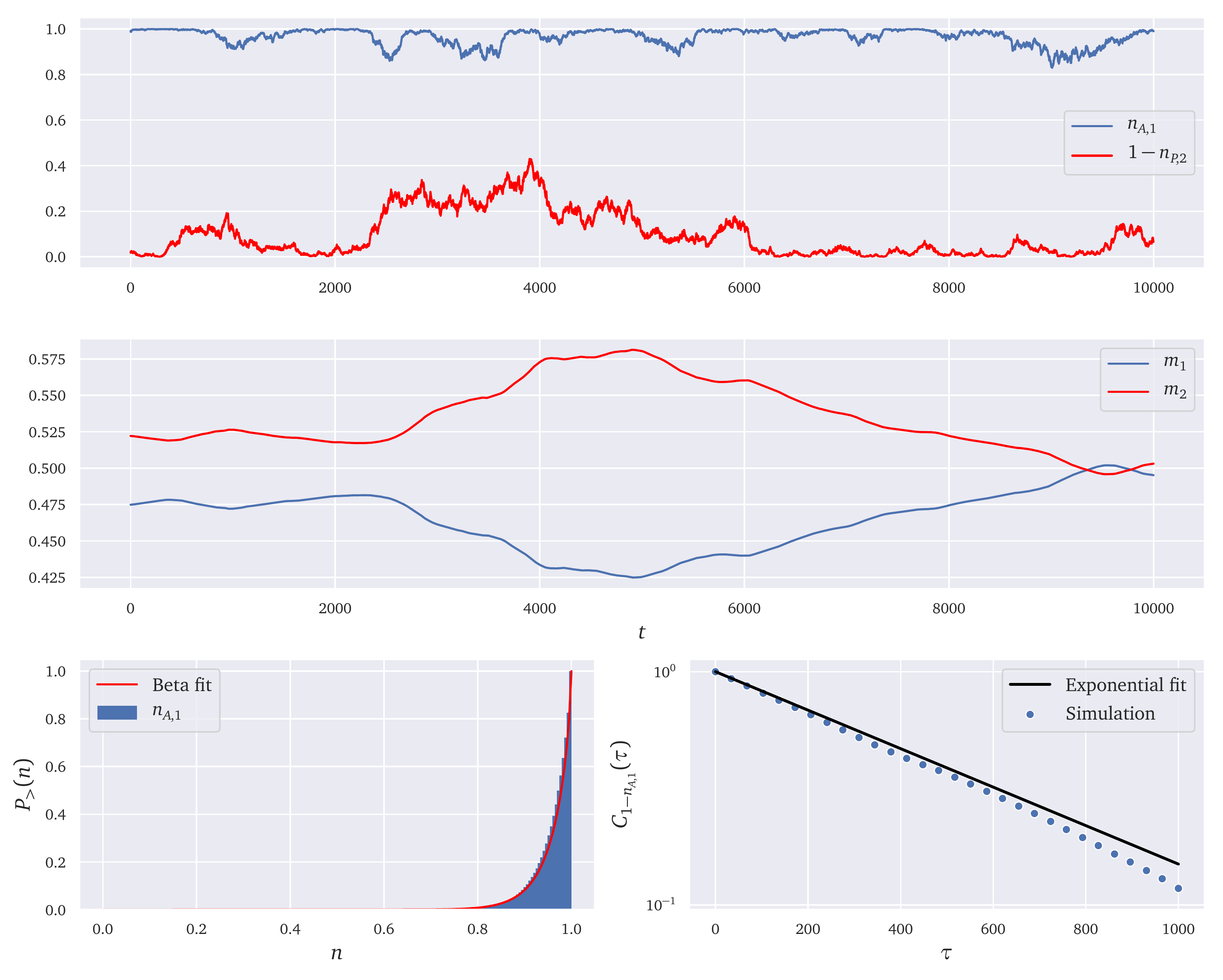}
  \caption{Numerical simulation of our model. We have chosen the different parameters to obtain the same stationary Beta distribution as that observed in Fig.~\ref{fig:cdf_data}. Note also the similarity of the Figure on the left with plot (a) in Figure~\ref{fig:fishing_zones_stats}. We have used the parameters $\varepsilon=0.69$, $\nu=10$, $\beta =5$, $C_N=1$, $C_{\varepsilon} = 26.5$ for Ancona and $15.7$ for Pescara and $\kappa=0.1$. The upper panel shows  two trajectories $n_{\mathrm{A},1}(t)$ and $n_{\mathrm{P},2}(t)$. The middle panel shows the fish populations $m_1(t)$ and $m_2(t)$. The bottom left panel shows the cumulative density function for $n_{\mathrm{A},1}$ along with a Beta distribution fit. The bottom right panel shows the empirical correlation function as defined by Eq.~\eqref{eq:corr_fct_model} along with an exponential fit. Note that the fish populations oscillate around the theoretical mean-field value $m=0.5$, and that large oscillations coincide with large collective movements of the fishermen in both areas.}
  \label{fig:simulations}
\end{figure}

Our model thus replicates successfully the observed distributions shown in Fig.~\ref{fig:cdf_data}, and captures the qualitative behaviour from Fig.~\ref{fig:fishing_zones_stats}. An example of numerical simulation of the model is provided in Fig.~\ref{fig:simulations}. For this Figure, we have set $\mu=1$ as it only amounts to a certain choice of the timescale, while picking a small value $\kappa=0.1$ to keep $f_2,f_1\approx 1$. We have then picked $\varepsilon$ and $C_{\varepsilon}$ such as to match the values for $\gamma_0$ and $\gamma_1$ from Fig.~\ref{fig:cdf_data}.

\subsection{Dynamics and correlation functions} 
\label{sub:dynamics_and_correlation_functions}

Given a Fokker-Planck equation of the form
\begin{equation}\label{eq:fp_example}
\partial_t f = -\partial_x\left[a(x,t)f\right] + \frac{1}{2}\partial^2_{xx}\left[b(x,t)^2f(x,t)\right],
\end{equation}
it is possible to show that it describes the probability density of a process determined by the following Ito stochastic differential equation:
\begin{equation}\label{eq:sde_example}
\frac{\dint x}{\dint t} = a(x,t)+b(x,t)\eta(t),
\end{equation}
where $\eta(t)$ is a Gaussian white noise of unit variance. The process of obtaining the Fokker-Planck equation from Eq.~\eqref{eq:sde_example} is described in detail in ~\cite{risken1996fokker}.

Taking the mean-field model defined above, it is straightforward to show that the variable $n_{\mathrm{A},1}$ follows a stochastic differential equation:
\begin{equation}\label{eq:sde}
\frac{\dint n_{\mathrm{A},1}}{\dint t} = \mu \left(\gamma_{\mathrm{A},0}- \left(\gamma_{\mathrm{A},0}+\gamma_{\mathrm{A},1}\right)n_{\mathrm{A},1}\right)+\sqrt{2\mu n_{\mathrm{A},1}\left(1-n_{\mathrm{A},1}\right)}\eta(t),
\end{equation}
with $\eta$ a Gaussian white noise of unit variance. This equation may be solved formally by integration, just as one would for a standard ordinary differential equation, to obtain

\begin{equation}\label{eq:sde_sol}
\begin{split}
n_{\mathrm{A},1}(t) = &\frac{\gamma_{\mathrm{A},0}}{\gamma_{\mathrm{A},0}+\gamma_{\mathrm{A},1}}e^{-\mu(\gamma_{\mathrm{A},0}+\gamma_{\mathrm{A},1})t}n_{\mathrm{A},1}(0)\\
&+ \int_{0}^t \dint s~e^{-\mu(\gamma_{\mathrm{A},0}+\gamma_{\mathrm{A},1})(t-s)}\sqrt{2\mu n_{\mathrm{A},1}(s)\left(1-n_{\mathrm{A},1}(s)\right)}\eta(s).
\end{split}
\end{equation}

From Eq.~\eqref{eq:sde}, we can show directly that the auto-correlation of $1-n_{\mathrm{A},1}$, defined as in Eq.~\eqref{eq:corr_function} and labelled $C_{1-n_{\mathrm{A},1}}(\tau)$, satisfies the following differential equation,
\begin{equation}\label{eq:corr_fct_ode}
\frac{\dint}{\dint \tau}\left[C_{1-n_{\mathrm{A},1}}(\tau)\right] = - \mu \left(\gamma_{\mathrm{A},0} + \gamma_{\mathrm{A},1}\right)C_{1-n_{\mathrm{A},1}}(\tau)
\end{equation}
which can easily be solved with the condition that $C_{1-n_{\mathrm{A},1}}(0)=1$, yielding
\begin{equation}\label{eq:corr_fct_model}
C_{1-n_{\mathrm{A},1}}(\tau) = \exp\left(- \mu \left(\gamma_{\mathrm{A},0} + \gamma_{\mathrm{A},1}\right)\tau\right),
\end{equation}
which is exactly what one sees from the data in Fig.~\ref{fig:fishing_zones_stats}~(d) and (e), provided one interprets the delta-peak at $\tau=0$ as the result of exogenous noise, e.g. weather conditions.

Indeed if one considers that the measured signal $\tilde{n}$ is, in fact, a noisy signal,
\begin{equation}\label{eq:noisy_measurement}
1-\tilde{n}(t)=(1-\sigma)(1-n(t)) + \sigma \xi(t),
\end{equation}
where $n(t)$ is the ``true'' process and $\xi(t)$ is a Gaussian white noise of unit variance, then one can show directly that the measured correlation function reads:
\begin{equation}\label{eq:noisy_corr_function}
C_{1-\tilde{n}}(\tau) = \frac{\sigma^2 }{\sigma^2 +(1-\sigma)^2}\delta\left(\tau\right)+\frac{(1-\sigma)^2}{\sigma^2 +(1-\sigma)^2 }C_{1-n}(\tau).
\end{equation}

\begin{figure}[tb]
  \centering
  \includegraphics[width=\textwidth]{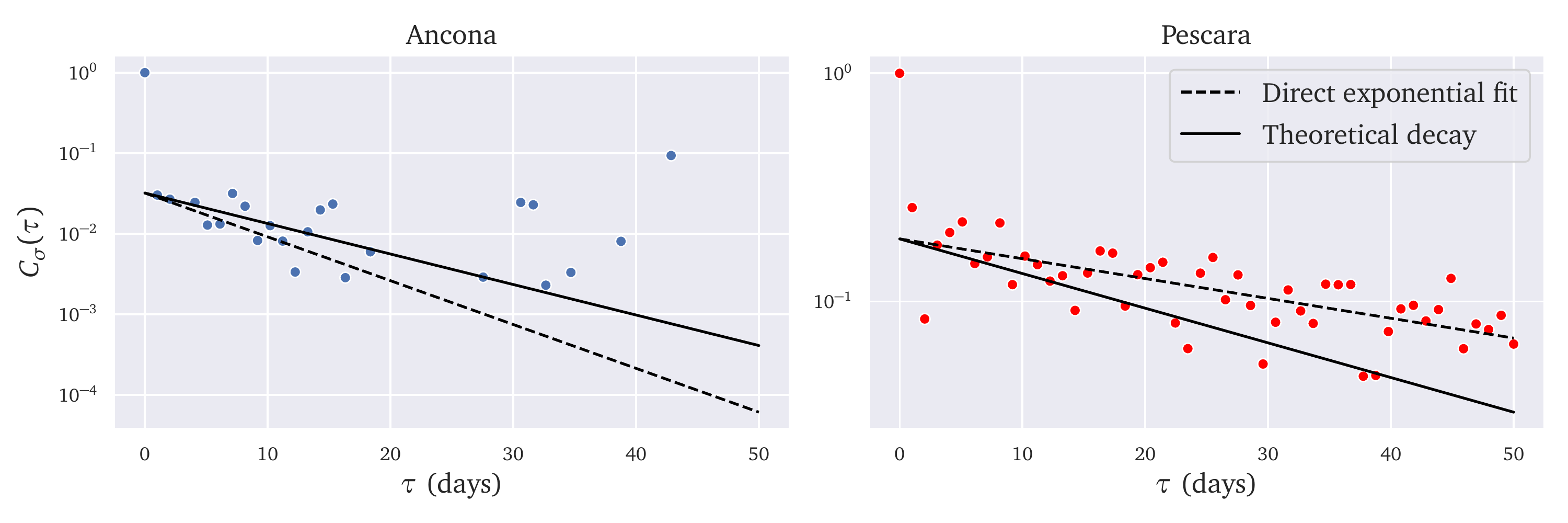}
  \caption{Empirical correlation function $C_{\sigma}(\tau)$ as defined in Eqs.~\eqref{eq:sigma2_body} and~\eqref{eq:sigma2_decay}. The solid black line is the theoretical prediction given the estimations of $\gamma_0$ and $\gamma_1$ from the empirical probability distribution in Fig.~\ref{fig:cdf_data} and from the subsequent estimation of $\mu$ using the exponential decay factor from Fig.~\ref{fig:fishing_zones_stats}. The reliable computation of $\sigma$ depends of course on the proper estimation of these parameters, and we expect them to be noisy. Nonetheless, the agreement with theory, especially in the case of Pescara, is rather good.}
  \label{fig:correlations2}
\end{figure}

We have checked that the correlation function given in Eq.~\eqref{eq:corr_fct_model} agrees with our numerical simulations as well; the results are shown in the bottom right panel in Fig.~\ref{fig:simulations}.  

The agreement is excellent both between the mean-field theory and the data, indicating that our model can correctly replicate the main dynamical features of real data from fishing dynamics.  Furthermore, one can deduce the value of $\mu$ from the values of $\gamma_0,\, \gamma_1$ and the decay factor in the exponential, that should match $\mu\left(\gamma_0+\gamma_1\right)$. Using this formula, we find $\mu=4.3\cdot 10^{-3}$ for Ancona and $\mu = 1.7\cdot 10^{-3}$ for Pescara. 
 
Note that the middle panel in Fig.~\ref{fig:simulations} shows very interesting dynamics, with the fish population oscillating about its average value $\avg{m_1}=\avg{m_2}=0.5$ through visible interactions with the amount of boats fishing in the two areas. Note for example that the fish population tends to increase when the boats leave an area, but this is detrimental to the fish in the other area. In the Figure, it is also visible that boats from Pescara arriving at $t\approx 7000$ arriving in the zone close to Ancona had a detrimental effect to the population $m_1$ of fish close to Ancona, and the population in that zone did not necessarily have the time to recover from that excess in the time allowed by the simulation.

More complicated correlation functions can also be computed, along the lines of \cite{moran2020schr}, although they are more prone to statistical noise. For example, using the techniques described in detail in ~\cite{moran2020schr}, one can show that the polynomial defined by:
\begin{equation}\label{eq:sigma2_body}
\sigma_{\mathrm{A}}(n_{\mathrm{A},1}) = n_{\mathrm{A},1}^2 - \frac{2 (\gamma_{\mathrm{A},0} + 1)}{\gamma_{\mathrm{A},0} + \gamma_{\mathrm{A},1} + 2} n_{\mathrm{A},1},
\end{equation}
has an autocorrelation function that is exponential, meaning that $C_{\sigma_A}(\tau)=\Cor\left(\sigma_{\mathrm{A}}\left(n_{\mathrm{A},1}(t+\tau)\right),\sigma_{\mathrm{A}}\left(n_{\mathrm{A},1}(t)\right)\right) $ verifies:
\begin{equation}\label{eq:sigma2_decay}
C_{\sigma_{\mathrm{A}}}(\tau) = \exp\left(- 2 \mu (1 + \gamma_{\mathrm{A},0} + \gamma_{\mathrm{A},0}) \, \tau\right),
\end{equation}
and the same can of course be transposed to the variables indexed by $\mathrm{P}$.

We have tested this prediction in Fig.~\ref{fig:correlations2}. This correlator is necessarily more affected by measurement noise, because it is of order two in the $n$ variables and because it depends on a reliable estimation of the $\gamma$ and $\mu$ variables. Considering these limitations, the theoretical prediction is rather satisfactory when compared with the data, especially in the case of Pescara.

The reader should note here that the correlator described in Eq.~\eqref{eq:sigma2_decay} is not a feature that is present in the data but a direct consequence of our model. The Figure~\ref{fig:correlations2} shows that the data reproduces the expected behaviour reasonably well, thereby showing that our model has a very strong coherence with the dynamics observed in the data. 

\begin{figure}[t!]
  \centering
  \includegraphics[width=\textwidth]{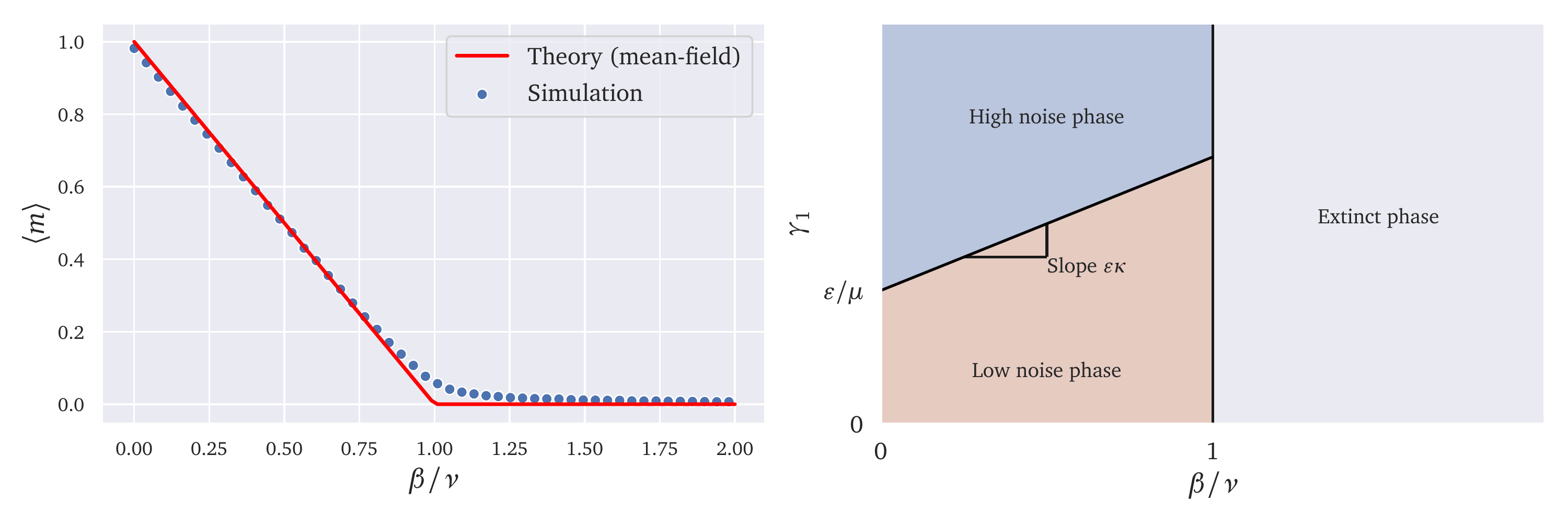}
  \caption{Left: Numercial results (same parameters as in Figs~\ref{fig:simulations} and \ref{fig:correlations2}). The simulation was run for $T=10^5$ steps and with $\nu=10$. Note that the convergence of the simulation to the mean-field results from Eq.~\eqref{eq:extinction_line_sym} improves as $T$ or $\nu$ grow larger. Right: Phase diagram of the model. Note that the high/low noise frontier line could be extended in the extinct phase; indeed, without further ingredients, in the extinct phase boats move between the two areas as in the non-extinct phase even though they are not able to fish anymore. Note that the linear frontier of slope $\varepsilon \kappa$ between the low and high-noise phases could be more complicated if a different function is picked for $f$. }
  \label{fig:extinction}
\end{figure}

\section{The symmetric limit} 
\label{sec:sym_limit}

In general, the fixed point equations defined at the beginning of Section~\ref{section:mean-field} linking the averages $\avg{m_i}$ with the averages $\avg{n_i}$ cannot be solved directly. Nonetheless, if one takes $C_N=1$ to have two fishing areas that are perfectly symmetric, then the equations simplify considerably as this immediately implies $f_1=f_2$, with  Eq.~\eqref{eq:avg_na1} becoming:
\begin{equation}\label{eq:avgn_symmetric}
\avg{n_{\mathrm{A},1}} = \avg{n_{\mathrm{P},2}} = \frac{C_{\varepsilon}}{C_{\varepsilon}+1}.
\end{equation}
One can then write Eq.~\eqref{eq:mean_field_avg} more explicitly, to obtain the following extinction line:
\begin{equation}\label{eq:extinction_line_sym}
\avg{m_1} = \avg{m_2} = \left\{ \begin{matrix}
1-\frac{\beta}{\nu} &\text{ if }& \beta<\nu\,\\
0 &\text{ if }&  \beta \geq \nu ,
\end{matrix}\right.
\end{equation}
which has the intuitive interpretation that the population within a given area goes extinct if the fishing rate is larger than the reproduction rate of fish in the area.  Figure~\ref{fig:extinction} shows that the agreement of numerical simulations with our mean-field analysis is excellent. One should note however that convergence may be slow when $\nu\to 0$, as this parameter controls the global fish population typical time-scale.

Note also that in this case, one can directly compute $f_1=f_2 = 1+ \frac{\kappa \beta}{\nu}$. The parameters in Eq.~\eqref{eq:gamma_params} simplify to yield:
\begin{equation}\label{eq:gamma_params_simplified}
\gamma_{0} = \frac{\tilde{\varepsilon}}{\mu}C_{\varepsilon},\quad \gamma_{1} = \frac{\tilde{\varepsilon}}{\mu},
\end{equation}
where we've dropped the $\mathrm{A}$ index as the parameters for both areas $\mathrm{A}$ and $\mathrm{P}$ are identical, and where we've set $\tilde{\varepsilon}:=\varepsilon\left(1+\frac{\kappa \beta}{\nu}\right)$.  

In this limit it is very clear that our mean-field model amounts to a modification of the original ant model \cite{kirman1993ants}, where the noise $\varepsilon$ is augmented because of the sensitivity of the fishermen to the local fish population by the factor given above, and where we have introduced an asymmetry between the two areas/food-sources through the parameter $C_{\varepsilon}$.

One would then typically expect to have always have $\gamma_0>1$ because of the strong preference for the fishing area closest to one's home port. However, if $\varepsilon$ or $\kappa$ are strong enough, one can have a crossover at $\gamma_1=1$, separating a regime where the boats are all frequently found to be close to their home port, corresponding to $\gamma=1$ and $n_{\mathrm{A},1}=1$ as the most probable value in the distribution, from a regime where there is mixing and the most probable value is $n_{\mathrm{A,1}}<1$, corresponding to $\gamma_1>1$.

\begin{figure}[tb]
  \centering
  \includegraphics[width=\textwidth]{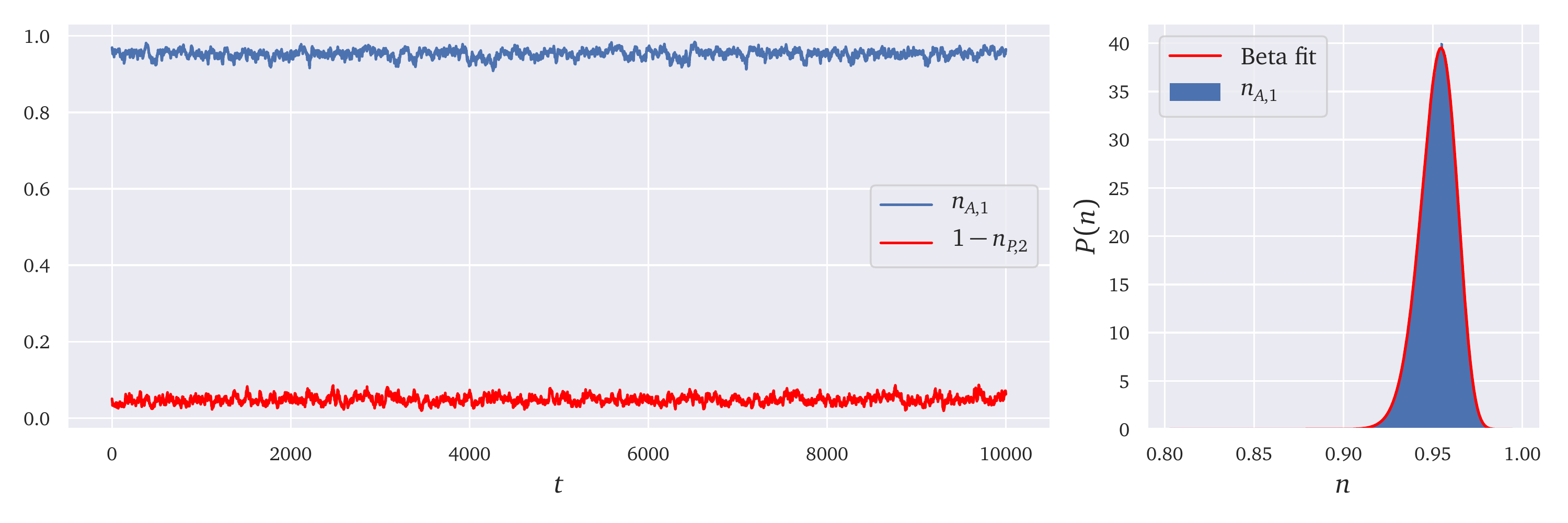}
  \caption{Numerical simulations in the case $\gamma_1>1$. The parameters are the same as that of Fig.~\ref{fig:simulations}, but with $\kappa=10$ instead. The Beta fit is compatible with  the predicted theoretical values from Eq.~\eqref{eq:gamma_params_simplified}. Note that, in contrast with  Fig.~\ref{fig:cdf_data}, we show the probability density  instead of the cumulative density function.}
  \label{fig:difference}
\end{figure}

The simulations displayed in Fig.~\ref{fig:simulations} correspond  to $\gamma_1<1$, the empirical data shown in Figs~\ref{fig:fishing_zones_stats} and~\ref{fig:cdf_data} has $\gamma_1<1$ for Ancona, and $\gamma_1\gtrsim 1$ for Pescara. As stated above, when $\gamma_1>1$ the behaviour is qualitatively different: instead of having the majority of the boats nearly always fish in their home area, with occasional ``jumps'' to go to the neighbouring zone, there is always a degree of ``mixing'', as at any given time there is always a fraction $\approx 1-\avg{n_{\mathrm{A},0}}$ of fishermen from Ancona fishing near Pescara. For the sake of completeness Fig.~\ref{fig:difference} displays a simulation of this case, with $\gamma_1$ well above $1$. This last regime is qualitatively very different to that with $\gamma_1<1$ and corresponding to Figs.~\ref{fig:fishing_zones_stats} (top panel) and~\ref{fig:simulations}, as stressed above: there are always close to $95\%$ of the fishermen from Ancona fishing near their home port with no large collective excursions to Pescara's area.

\section{Discussion} 
\label{sec:exploitation_of_finite_resources}

\begin{figure}[tb]
  \centering
  \includegraphics[width=\textwidth]{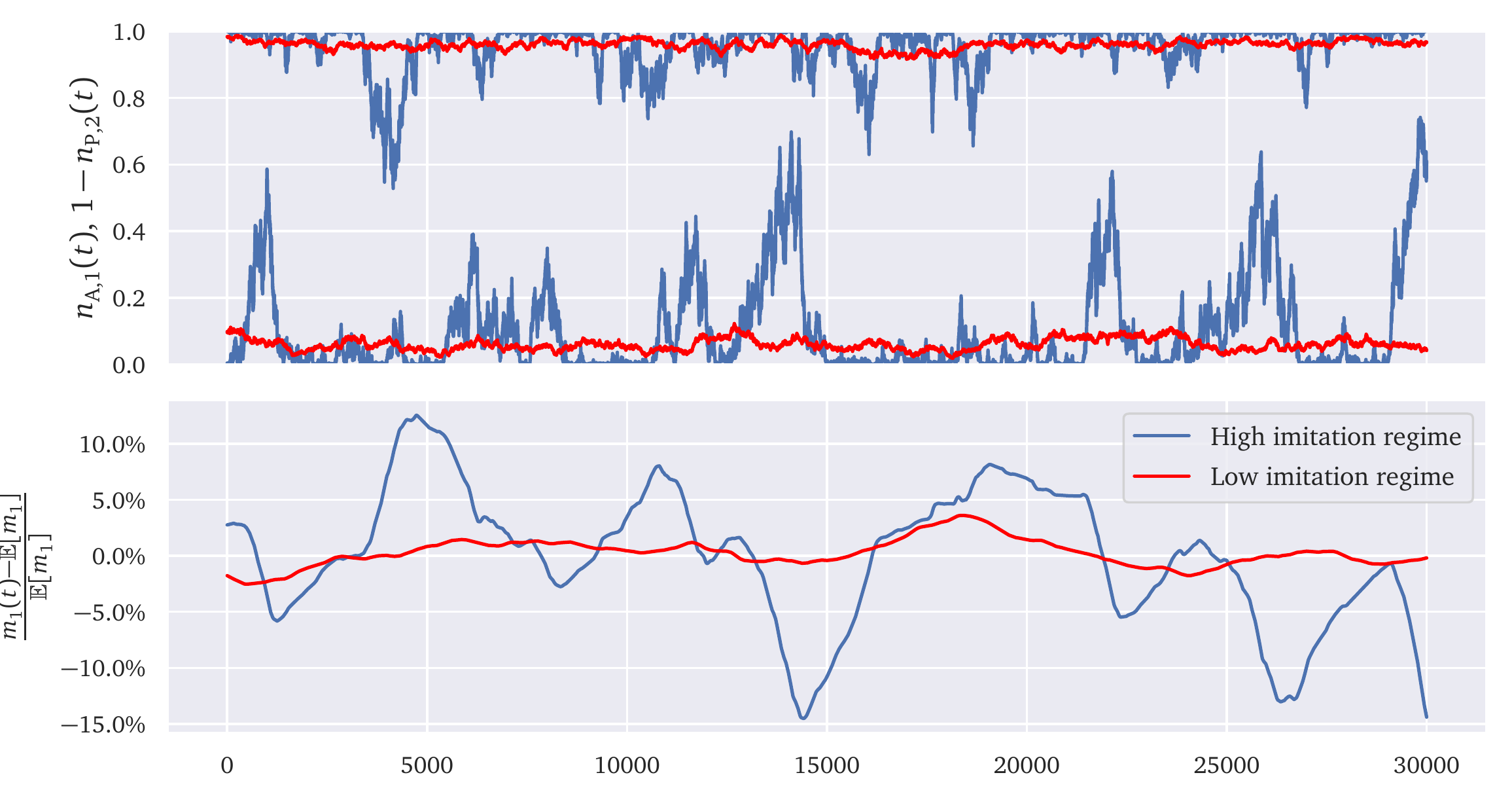}
  \caption{A more detailed comparison between the two different regimes of low and high imitation. The simulations correspond to two ports with the same number of boats ($C_N=1$) and with $C_{\varepsilon}\approx 20$, with $\varepsilon=1$ for both. The low imitation regime (in red) corresponds to $\mu=0.1$ while the high imitation regime (in blue) has $\mu=5$. The fractions $n_{\mathrm{A},1}$ and $1-n_{\mathrm{P},2}$ are represented on the top plot, the former being the one closest to $1$ and the latter the one closest to $0$, and their sum is proportional to the total number of fishing boats in the first fishing area. The bottom curve shows the fish population in that area, and one sees that large fluctuations of the number of boats there translate into large population fluctuations; the same effect is seen on the total catch of boats based in port A. It is therefore clear that the large, imitation-induced switches of the fishing boats in the large $\mu$ regime cause the fish population to fluctuate wildly.}
  \label{fig:vol_comparison}
\end{figure}

The interesting feature of this model is that it shows explicitly the effect of herding on finite resources, which is the population of fish in this case. The dynamics at play are very easy to understand intuitively: if imitation is too strong, then boats will have a tendency to pool together in one area, which will lead the fish population there to decrease until it becomes very probable that a boat decides to leave and trigger an ``avalanche'' of other boats that will follow suit; the second area will then also be overfished while the first replenishes, until another large switch happens. This causes the fish population in the two areas, and therefore also the catch that can be brought back to each port, to be very volatile in the high imitation regime. This behaviour can be seen easily in Figure~\ref{fig:vol_comparison}.

Again this behaviour is reminiscent of the one often seen in financial markets, as described in detail in~\cite{alfarano005,Alfarano2007,alfarano_2008,sano2015,chiarella1992dynamics} and ~\cite{sano2015}. In those models, the agents' tendency to imitate each other causes large swings of opinion and pushes them to over- or under-value the price of an asset. Once the price has been too far off from a ``fundamental'' value, correcting mechanisms kick in that correct this, but the imitative behaviour causes this correction to overshoot until the process repeats again. As stated before, these models are not only allegorical, as recent empirical evidence shows that this behaviour does indeed appear in real data~\cite{Majewski2018,bouchaud2017black}, and can cause instabilities and bubbles to develop in financial markets.

\begin{figure}[tb]
  \centering
  \includegraphics[width=\textwidth]{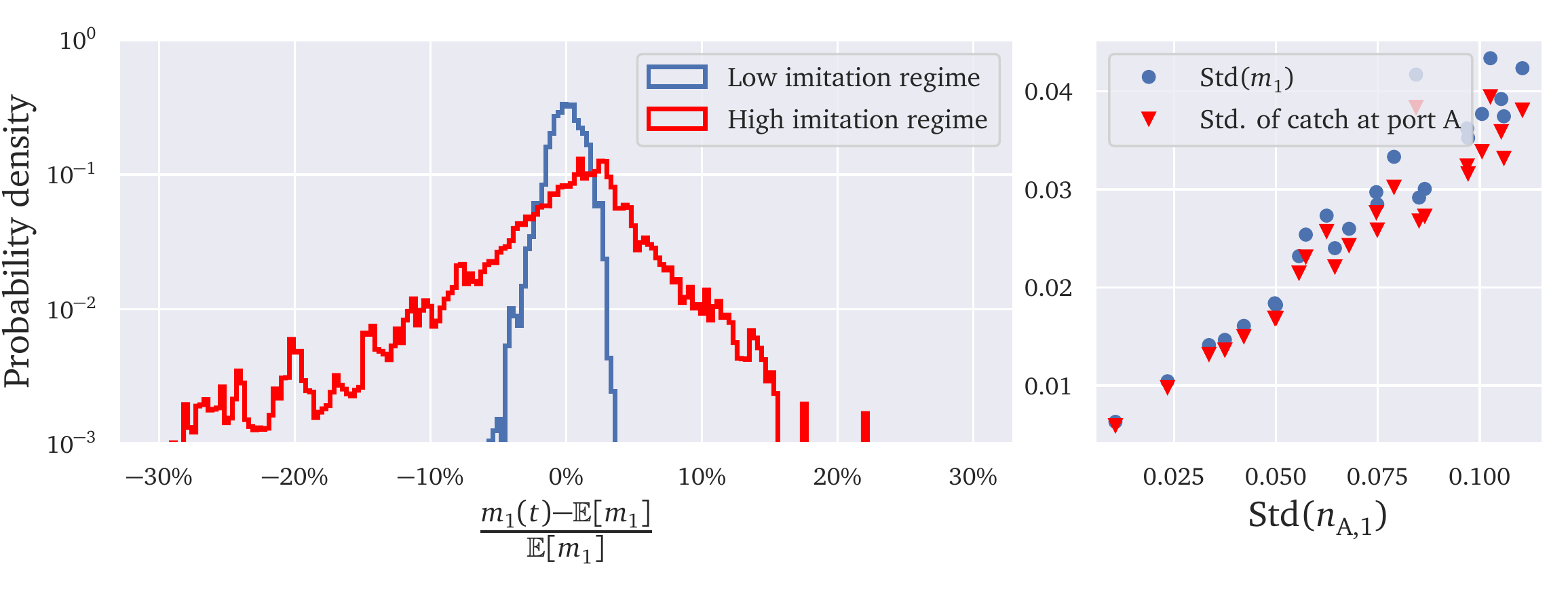}
  \caption{Left: distribution of the relative fluctuations of the fish population in zone $1$ with respect to its average. The high imitation regime again corresponds to $\mu=5$ and the low imitation to $\mu=0.1$. It is clear that the high imitation regime can cause extremely large fluctuations to appear, some causing the fish population to be at times almost $30\%$ lower than its average value. Right: standard deviations of $m_1$ and of the catch that is brought to port by the boats at port $\mathrm{A}$, proportional to $n_{\mathrm{A},1}(t)m_1(t)+(1-n_{\mathrm{A},1}(t))m_2(t)$}. Note that the standard deviation of $n_{\mathrm{A},1}$ increases with $\mu$ in the high-imitation regime, as seen in Eq.~\eqref{eq:mean_var}. Also, because $n_{\mathrm{A},1}(t)$ is often closest to $1$ because of the asymmetry of the fishing areas it follows that these two quantities closely match each other. Therefore the high imitation regime is not only detrimental to the stability of the fishes' ecosystem, since they often prone to approach depletion, but also to the overall efficiency of the boats' fishing in terms of yield.
  \label{fig:hist_catch_scatter}
\end{figure}

Similarly, we see here continued cycles of overfishing, but the fact that we are dealing with finite resources makes the situation overall more dangerous.  If the imitation is strong enough, so that boats persist in going to an area that has approached depletion because they assume imitating their peers is the best way to proceed, then it is possible that in one of these cycles an area is effectively depleted and that no replenishment is possible.  The boats will then leave it and go to the other area, that is incapable of sustaining them all and will inevitably be depleted too. This is a case where the strategy of imitation seems rational from the standpoint of the solipsistic individual unconcerned with the systemic impact of his actions, and impervious to the fact that his choice will not only add further stress to a fishing area but will also increase the odds that more agents do as he. The low imitation phase is closer to the ``optimal'' behaviour both for fishermen and for fishes: as seen in Figure~\ref{fig:vol_comparison}, this regime suppresses the large oscillations in fish population and allows it to quickly reach a stationary value that corresponds to its average.

This has natural consequences for the way we should strive to manage natural resources. One possibility would be that the best overall catch would be obtained by simply having all the boats from each port fishing in their own area, while also ensuring that their fishing does not deplete the area -- one could imagine, for example, imposing regulation to lower the $\beta$ and therefore the amount of fish caught by each boat. But this would not be the case if, for example there are more boats registered in one of the two areas. It should also be noted that there is a ``loss'', or cost, for a boat fishing in an area different from its home area. So, if cross fishing is occurring the situation could be improved by two boats switching areas. But there is no obvious mechanism that would ensure that this happens. Furthermore, there is certainly some noise in the information that boats receive which will lead to some movement even when it does not seem to be objectively justified.


\section{Conclusion}

In this paper, we analysed the empirical distribution of the locations of fishing vessels in the two areas near to Ancona and Pescara. By detecting to which port a vessel belongs, we computed the fraction of fishermen fishing in their home zone and looked at their statistical properties. We found that the empirical distribution functions are well approximated by asymmetric Beta distributions, and their auto-correlations by exponentials. Inspired by such evidence, we extended Kirman and Föllmer's ants recruitment model to finite and asymmetric resources. We performed a numerical and theoretical analysis in the mean field approximation, and showed that the auto-correlations and the stationary distribution of the fraction of fishermen  appear to be respectively exponential and Beta distributed. Finally, we provided the phase diagram that separates a low and high herding phase, as well as a fish extinction phase.

We have further tested our dynamics by looking at higher order correlations that can be empirically computed. This signal appears to be very noisy and of low intensity but consistent with an exponential decay, with a time-scale compatible with that predicted by our model. Given the results that we have described, we are quite confident in our minimal model since it is able to reproduce surprisingly well the generic stylized facts within a limited though behaviourally sound framework. In particular, we have shown in Appendix~\ref{appendix:multi_zones} that while a multi-zone model (with more than two zones) would possibility be more realistic, the results for our two-zone model can be seen as the result of the aggregation of several zones, providing solid micro-foundations to our approach and justifying our looking at two aggregated zones for empirical analysis.

In the spirit of ~\cite{moran2020schr}, we have introduced in this paper a set of analytical techniques, namely deriving and solving a Fokker-Planck/Kolmogorov forward equation, that allow to gain further insight into the collective dynamics of interacting agents. In contrast to past work exploring similar models, we have also introduced a study of the different correlators of the model and have shown how stochastic calculus techniques can give a clearer view of the dynamics of the model. This is, along with the mean-field treatment of Section~\ref{section:mean-field}, the main methodological contribution of this paper.

Finally, our analysis of the behaviour of the boats showed that strong imitation can have direct consequences on the overall stability of the ecosystem and those whose livelihood depends on it. The fish population dynamics inherits the volatile behaviour of the boats, and the possibility of extinction looms if this volatility is too strong. Thus, the fishermen, acting in their own self-interest and absent any type of coordination except for their imitative behaviour, can cause their livelihood to disappear. In our particular case, of course, this has not happened, which can either be attributed to the fact that the incentive to imitate has not been sufficiently strong or that the number of boats and/or the amount they are able to catch is not yet excessive.  However, even a limited amount of imitation has a negative effect on the overall yield of fish. 

This raises two interesting questions. Firstly, to what extent were areas which were heavily depleted in the past the victims of strong herding behaviour as more vessels moved into them rather than to the existing fleets fishing in those areas extracting too much? Secondly, how effective are quotas in attenuating this phenomenon? 

\section*{Acknowledgements}

We thank Jean-Philippe Bouchaud, Alexandre Darmon, Mauro Gallegati and Gianfranco Giulioni for fruitful discussions and help in interpreting the data.
This research was conducted within the \emph{Econophysics \& Complex Systems Research Chair}, under the aegis of the Fondation du Risque, the Fondation de l'\'{E}cole polytechnique, the \'{E}cole polytechnique and Capital Fund Management, and within the \textit{New Approaches to Economic Challenges} (NAEC) OECD programme on complex systems.

\clearpage

\bibliographystyle{apalike}
\bibliography{biblio.bib}

\appendix

\clearpage

\section{Full dynamical solution}
\label{ap:dynamics}

The goal of this section is to sketch a full dynamical solution for the dynamics of Eq.~\eqref{eq:sde}. We drop the indices $\mathrm{A}$ or $\mathrm{P}$ for clarity, obtaining:
\begin{equation}\label{eq:sde_general}
\frac{\dint n}{\dint t} = \mu \left(\gamma_0- \left(\gamma_0+\gamma_1\right)n\right)+\sqrt{2\mu n (1-n)}\eta(t),
\end{equation}
a stochastic differential equation that corresponds to the following Fokker-Planck equation \cite{risken1996fokker}:
\begin{equation}\label{eq:fp_general}
\partial_t \rho = \mu \partial_{nn} \left( n(1-n) \rho \right) - \mu \partial_n \left( \left(\gamma_0- \left(\gamma_0+\gamma_1\right)n\right) \rho \right),
\end{equation}
with reflecting boundary conditions in $n=0$ and $n=1$.

As in ~\cite{moran2020schr}, one can ``diagonalize'' this equation, writing it as: 
\begin{equation}
   \partial_t \rho = \mathcal{A}\rho,
\end{equation} 
with $\mathcal{A}$ a Fokker-Planck operator that gives the right-hand side of Eq.~\eqref{eq:fp_general} when applied to $\rho$. It is in principle possible to apply the same techniques as in ~\cite{moran2020schr} to obtain a Schr\"odinger's equation for an alternative function $\Psi$, that one could then use to compute $\rho$ explicitly. 

However, one can also solve the eigenvalue problem $
  \mathcal{A}\rho_{\mathcal{E}} = \mathcal{E}\rho_{\mathcal{E}},
$
so that the general solution for $\rho$ reads:
\begin{equation}\label{eq:gen_sol}
  \rho(n,t) = \sum_{\mathcal{E}}\lambda_{\mathcal{E}}\rho_{\mathcal{E}}(n)e^{-\mathcal{E}t}.
\end{equation}
In this setting, $\mathcal{E}$ and $\rho_{\mathcal{E}}$ are respectively the eigenvalues and eigenvectors of the operator $\mathcal{A}$. These eigenvectors should also be normalized so that the integral of $\rho$ is equal to $1$.

The problem therefore translates into finding functions $\rho_{\mathcal{E}}$ and numbers (or ``energies'') $\mathcal{E}$   that satisfy:
\begin{subequations}
\begin{align}
&\mu \mathcal{E} \rho_\mathcal{E}  =  \mu \partial_{nn} \left( n(1-n) \rho_\mathcal{E} \right) - \mu \partial_n \left( \left(\gamma_0- \left(\gamma_0+\gamma_1\right)n\right) \rho_\mathcal{E} \right) \label{eq:fp_eigvect}
\\
&J_\mathcal{E} (0)  =  J_\mathcal{E} (1) = 0
\\
&\textstyle\int_0^1\dint n~ \rho_\mathcal{E}(n)    <  \infty,
\end{align}
\end{subequations}
with $ J_\mathcal{E}(n) = \mu \partial_n \left( n(1-n) \rho_\mathcal{E} \right) - \mu  \left(\gamma_0- \left(\gamma_0+\gamma_1\right)n\right) \rho_\mathcal{E} $.

In order to solve Eq.~\eqref{eq:fp_eigvect}, we rewrite it as:
\begin{equation}\label{eq:eq_diff_hypergeo}
 n(1-n) \rho_\mathcal{E}'' + \left(2 - \gamma_0 - \left(4 - \gamma_0 - \gamma_1  \right) n \right)  \rho_\mathcal{E}' - \left(2 + \mathcal{E} - \gamma_0 - \gamma_1 \right) \rho_\mathcal{E} = 0
\end{equation}
The solutions of this differential equation are given in terms of the hypergeometric function:
\begin{equation}
  _2F_1(a,b;c;n) = \sum_k \frac{(a)_k (b)_k }{(c)_k} \frac{n^k}{k!},\quad (a)_k = \prod_{i=0}^{k-1} (a+i).
\end{equation}
Here the two linear independent solutions well defined around zero, see~\cite{abramowitz1948handbook}, are $ _2F_1(a,b;2 - \gamma_0;n) $ and $n^{\gamma_0 - 1} _2F_1(a + \gamma_0 - 1,b + \gamma_0 - 1;\gamma_0;n)$, where $a,b$ are the solutions of:
\begin{subequations}
\begin{align}
a + b & =  3 - \gamma_0  - \gamma_1 \\
ab & =  2 + \mathcal{E} - \gamma_0 - \gamma_1.
\end{align}
\end{subequations}

Only the second solution cited above verifies the boundary condition at $n=0$. Applying then an Euler transformation\footnote{The Euler transformation for the hypergeometric function states that $_2F_1(a,b;c;n)  = (1-n)^{c-a-b} _2F_1(c-a,c-b;c;n) $.} on this solution leads to:
\begin{equation}
    \rho_\mathcal{E}(n) = C_\mathcal{E} n^{\gamma_0 - 1} (1-n)^{\gamma_1 - 1} \, _2F_1(1 - a, 1- b;\gamma_0;n),
\end{equation}
with $C_\mathcal{E} $ a constant. This solution is well defined at $n=1$ and  also verifies the boundary condition.  To check the integrability condition one can compute explicitly
\begin{equation}
    \int_0^1 \dint n~\rho_\mathcal{E}(n) = C_\mathcal{E} \sum_k \frac{(1-a)_k (1-b)_k \Gamma(\gamma_0 + k) \Gamma(\gamma_1) }{(\gamma_0)_k  \Gamma(\gamma_0 + \gamma_1 + k) k!},
\end{equation}
with $\Gamma$ the Gamma function. If $1-a$ is a non-negative integer\footnote{As the hypergeometric function is symmetric with respect to its two first arguments, we restrict our analysis to the first one only.} all terms in the series are non-zero. Using then $(x)_k \propto \Gamma(x+k)$ together with the Stirling formula $ \Gamma(x+1) \approx \sqrt{2 \pi} x^{x + 1/2} e^{-x}$ for $x \gg 1$, we find that the general term of the series converges to a constant when $k \to + \infty$ and therefore that $ \int_0^1\dint n~ \rho_\mathcal{E}(n)  = + \infty$.
Therefore, the condition that the functions $\rho_{\mathcal{E}}$ have a finite integral implies that there exists a positive integer $k$ such that $ 1 - a = -k$, and so also that $ b = 2 - k - \gamma_0  - \gamma_1 $ and $\mathcal{E} = - k (\gamma_0 + \gamma_1 + k - 1)$. Since the numbers $\mathcal{E}$ are discrete, and are indexed by $k$, we write $\rho_{k} := \rho_{\mathcal{E}_k}$.

In conclusion, the eigenvectors $\rho_k$ and eigenvalues $\mathcal{E}_k$ are discrete and given by:
\begin{eqnarray}
    \mathcal{E}_k & = & - k (\gamma_0 + \gamma_1 + k - 1) 
    \\
    \rho_k(n) & = & C_k n^{\gamma_0 - 1} (1-n)^{\gamma_1 - 1} \, _2F_1(-k, \gamma_0 + \gamma_1 + k - 1;\gamma_0;n),
\end{eqnarray}
which allows then for a solution of the form given in Eq.~\eqref{eq:gen_sol}.

There only remains to find the coefficients $\lambda_{\mathcal{E}}$ that depend on the initial condition. This can be done by transforming the Fokker-Planck equation into a Schr\"odinger's equation as in ~\cite{moran2020schr}, noticing that the solutions to said Schr\"odinger equation can be found in terms of the eigenvalues and eigenvectors $\rho_k$, and one can therefore find the coefficients $\lambda_{\mathcal{E}}$ by projecting the initial condition onto the orthogonal set of eigenvectors of the Schr\"odinger operators, see the Appendices in ~\cite{moran2020schr} for a detailed technical explanation.

\section{A symmetric multi-zones extension}\label{appendix:multi_zones}

Here we present a very natural extension of our model to the general case of $M$ symmetric zones with finite resources. Without loss of generality we set $C_{\varepsilon}=1$ to have lighter notations, but this does not change our main message. We also introduce the vector notations $ \mathbf{n}(t)=\left(n_1(t),\ldots,n_M(t)\right)$ and $ \mathbf{m}(t)=\left(m_1(t)\ldots,m_M(t)\right)$, where the index accounts for the zone, and call $p_{j \to i} (\mathbf{n} (t), \mathbf{m} (t))$ the infinitesimal probability that an agent initially present in zone $j$ at time $t$ moves to zone $i$ at $t+dt$. It follows that the evolution of $\mathbf{n}$ and $\mathbf{m}$ is given by:
\begin{eqnarray}\label{eq:multizones_dynamic}
 \dint m_i (t) & = & m_i (t) ( \nu (1 - m_i (t)) - \beta n_i (t)) \dint t \\
 p_{j \to i} (\mathbf{n} (t), \mathbf{m} (t)) & = &  \frac{n_j (t) N}{M - 1} \left[ \varepsilon  f \left(m_j (t)\right) \right] + \mu N^2 n_i (t) n_j (t) .
\end{eqnarray}
Introducing for simplicity $h(n,m) = m ( \nu (1 - m) - \beta n) $, the joint density $\rho$ of the variables $ (\mathbf{n} (t), \mathbf{m} (t))$ evolves according to the following Fokker-Planck equation:
\begin{equation}\label{eq:fk_multizones}
\begin{split}
    \partial_t \rho (\mathbf{n}, \mathbf{m}) = & - \sum_i \partial_{m_i} \left(h(n_i,m_i) \rho \right) + \sum_{i \neq j} \left(\partial_{n_i} - \partial_{n_j} \right) \left( \frac{\varepsilon f \left(m_j \right)}{M-1} n_j \rho \right) \\
    & + \mu \sum_{i \neq j} \left(\partial_{n_i n_i} - \partial_{n_j n_i} \right) \left( n_i n_j \rho \right).
\end{split}
\end{equation}

Owing to the symmetry of the problem, one may generalize the argument used in Section~\ref{section:mean-field} to obtain the stationary averages:
\begin{equation}\label{eq:stationarymean}
 \avg{n_i} = \frac{1}{M} , \quad
 \avg{m_i} = \left(1 - \frac{\beta}{M \nu}\right)_+ 
\end{equation}
with $(x)_+$ the positive part of $x$. This again shows the existence of an extinction regime whenever $\beta = M\nu$. In what follows we assume that $\beta / \nu < 1/M$, to study the behaviour of the system outside of extinction.

When $\kappa = 0$ the density of $\boldsymbol{n}$ is a Dirichlet distribution with all parameters equal to $\left( \epsilon /\delta \right)$, namely:
\begin{equation}
    \rho_{\boldsymbol{n}}(n_1, \ldots, n_M) = \left( \prod_{i = 1}^M n_i^{\epsilon /\mu} \right) \mathbf{1}_{\left\{ \sum_{i = 1}^M n_i = 1 \right\}}.
\end{equation}
As argued previously, whenever the noise-level is coupled to the fish population with $\kappa>0$, we postulate that the solution can be approximated by a Dirichlet distribution with all parameters set to $\left( \tilde{\varepsilon} /\mu\right)$ with $\tilde{\epsilon} = f(1 - \frac{\beta}{M \nu})\epsilon $. 

The Dirichlet distribution has one key property: $ \sum_{i \geq k} n_i$ follows a Beta distribution with parameters $ \left( k \tilde{\varepsilon} /\mu, (M-k) \tilde{\varepsilon} /\mu\right)$, corresponding to the stationary state of our two-zone model. We have also checked that the mean-field approximation of Eq.~\ref{eq:multizones_dynamic} follows the same type of property: the variable $  \sum_{i \geq k} n_i(t)$ is ruled by the mean-field approximation of Eq.~\eqref{eq:fokker_planck}. This result gives solid micro-foundations to our approach, and justifies our looking at two aggregated zones for empirical analysis. This likely contributes to the very good agreement found between empirical results and our model.

\end{document}